\documentclass[12pt]{article}
\usepackage[top=1in, bottom=1in, left=1in, right=1in]{geometry}

\usepackage{graphicx} 
\usepackage{graphics}
\usepackage{fancybox}
\usepackage{amsmath,latexsym,psfrag,epsf,epsfig,amssymb,rotating}
\usepackage{color}
\usepackage{multirow}
\usepackage{url}
\usepackage{natbib}
\usepackage{fullpage}
\usepackage{subfigure}
\usepackage{bm}

\usepackage[
            colorlinks]{hyperref}  
\hypersetup{linkcolor=blue,
           citecolor=blue,
           filecolor=black,
           urlcolor=blue,
           bookmarks=true
           }
\hypersetup{
    bookmarks=true,         % show bookmarks bar?
    unicode=false,          % non-Latin characters in Acrobat?s bookmarks
    pdftoolbar=true,        % show Acrobat?s toolbar?
    pdfmenubar=true,        % show Acrobat?s menu?
    pdffitwindow=false,     % window fit to page when opened
    pdfauthor={Alberto Caimo},     % author
    pdfnewwindow=true,      % links in new window
    colorlinks=true,       % false: boxed links; true: colored links
    linkcolor=blue,          % color of internal links
    citecolor=blue,        % color of links to bibliography
    filecolor=blue,      % color of file links
    urlcolor=blue           % color of external links
}
\newcommand{\pref}[1]{%
    \ref{#1} \ifnum\count0=\pageref{#1}\relax%
    \else (page \pageref{#1})\fi}

\newcommand{\eref}[1]{%
        \ref{#1}\ifnum\count0=\pageref{#1}\relax%
        \else {, p.\pageref{#1}}\fi}

\newcommand{\comment}[1]{}

\newenvironment{algorithm}{\vspace{5 mm}\sc}{\vspace{5 mm}}
\newlength{\labwidth}
\newcommand{\step}[1]{%
    \settowidth{\labwidth}{#1\ }%
    \par\noindent%
    \global\hangindent\labwidth {#1}%
    \hbox{ }%
    }%

\newcommand{\bftheta}{\boldsymbol{\theta}}

\newcommand{\bfSigma}{\boldsymbol{\Sigma}}

\newcommand{\bfmu}{\boldsymbol{\mu}}

\def\bfI{\textbf{\em I}}

\def\bfu{\textbf{\em u}}

\def\bfx{\textbf{\em x}}

\def\bfy{\textbf{\em y}}
\def\bfY{\textbf{\em Y}}

\newcommand{\EE}{{\mathbb{E}}}

\begin{document}

%\title{Bayesian model selection for\\ 
%       exponential random graph models}
% \author{Alberto Caimo \& Nial Friel\\
%         Clique Research Cluster \\
%         Complex and Adaptive Systems Laboratory \\
%         School of Mathematical Sciences,\\
%         University College Dublin, Ireland\\
%         \normalsize\texttt{\{alberto.caimo,nial.friel\}@ucd.ie}}
% \date{\today}
% \maketitle

\begin{center}

{\Large \bfseries Bayesian model selection for exponential random graph models} \vspace{3 mm}
 
 % {\large N. Friel\footnotetext[1]{\sl{{\it{Address for correspondence}}:
 % School of Mathematical Sciences, University College Dublin.\\
 % Email: nial.friel@ucd.ie}}$^{1\star\ddag}$, A. Caimo$^\dagger$ } \\[0.5cm]
 {\large A. Caimo$^\dagger$, N. Friel$^{\star\ddag}$ } \\[0.5cm]
 
 {\it $^\dagger${National Centre for Geocomputation, National University of Ireland, Maynooth, Ireland}\\
 $^\star${Clique Research Cluster, Complex and Adaptive Systems Laboratory, University College Dublin, Ireland} \\
 $^\ddag${School of Mathematical Sciences, University College Dublin, Ireland} }
 
\vspace{3 mm}
{\large \today}
 \vspace{1 mm} 

\vspace{20 mm}

\end{center}

% 
% \ifpdf
%     \graphicspath{{pictures/}}
% \else
%     \graphicspath{{pictures/EPS/}}
% \fi

\begin{abstract}
Exponential random graph models are a class of widely used exponential family models for social networks. The topological structure of an observed network is modelled by the relative prevalence of a set of local sub-graph configurations termed network statistics. One of the key tasks in the application of these models is which network statistics to include in the model. This can be thought of as statistical model selection problem. This is a very challenging problem---the posterior distribution for each model is often termed ``doubly intractable'' since computation of the likelihood is rarely available, but also, the evidence of the posterior is, as usual, intractable. The contribution of this paper is the development of a fully Bayesian model selection method based on a reversible jump Markov chain Monte Carlo algorithm extension of \cite{cai:fri11} which estimates the posterior probability for each competing model.
\end{abstract}

\section{Introduction}

In recent years, there has been a growing interest in the analysis of network data. Network models have been successfully applied to many different research areas. We refer to \cite{kol09} for an general overview of the statistical models and methods for networks.

Many probability models have been proposed in order to summarise the general structure of networks by utilising their local 
topological properties: the Erd\"{o}s-R\'{e}nyi random graph model \citep{erd:ren59} in which edges are considered Bernoulli 
independent and identically distributed random variables; the $p_1$ model \citep{hol:lei81} where dyads are assumed independent, 
and its random effects variant the $p_2$ model \citep{van:sni:zij04}; and the Markov random graph model \citep{fra:str86} where 
each pair of edges is conditionally dependent given the rest of the graph. 

\begin{figure}[htp]
\centering
\includegraphics[scale=0.7]{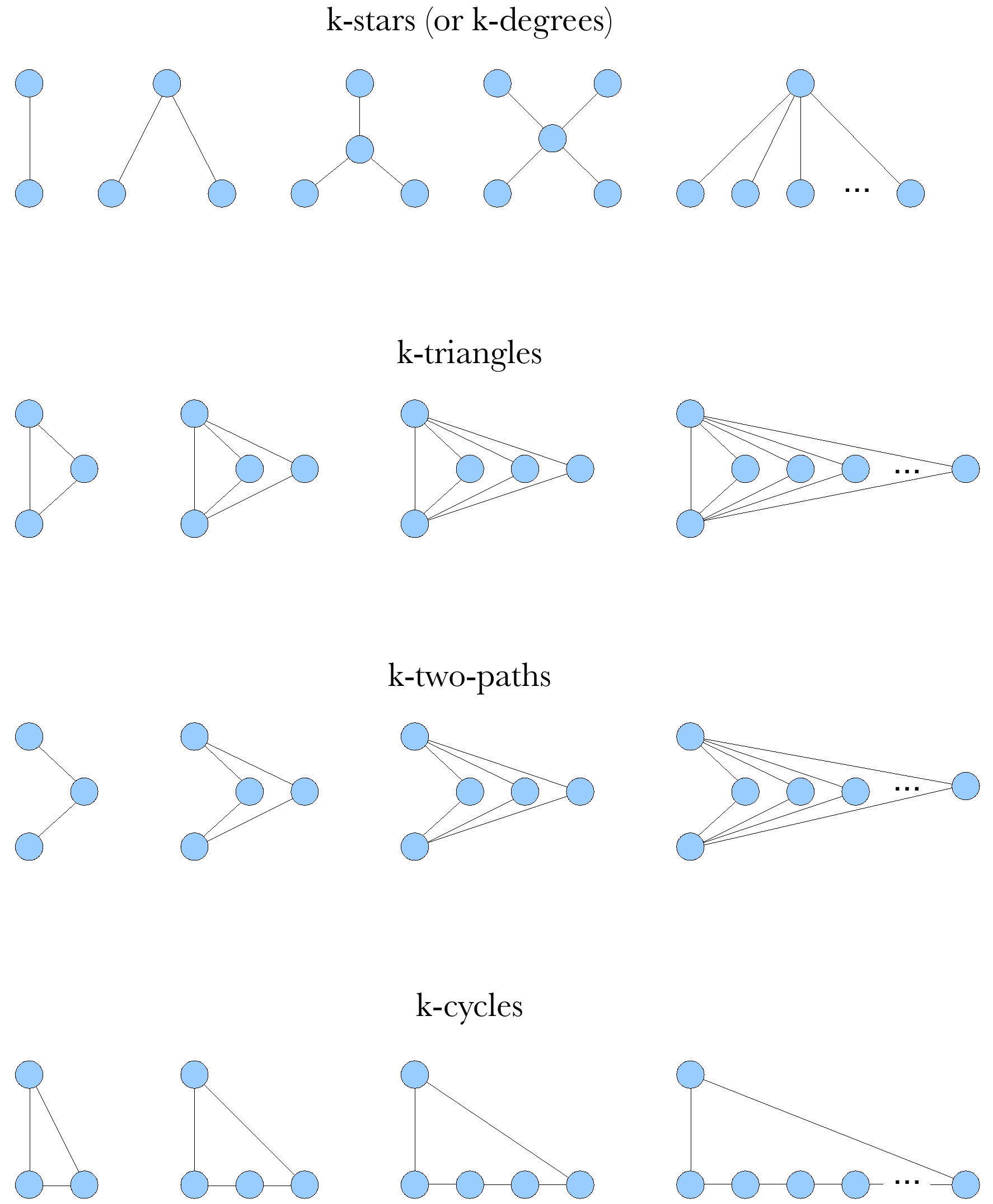}
\caption{Some of the most used sub-graph configurations for undirected graphs (analogous directed versions can be used for digraphs).}
\label{fig:specs}
\end{figure}

Exponential random graph models (see \cite{was:pat96,rob:sni:wan:han:pat07}) represent a generalisation of the latter model and 
have been designed to be a powerful and flexible family of statistical models for networks which allows us to model network 
topologies without requiring any independence assumption between dyads (pairs of nodes). These models have been utilized 
extensively in the social science literature since they allow to statistically account for the complexity inherent in many 
network data. The basic assumption of these models is that the topological structure in an observed network $\bfy$ can be explained 
by the relative prevalence of a set of overlapping sub-graph configurations $s(\bfy)$ also called graph or network statistics 
(see Figure~\ref{fig:specs}).

Formally a random network $\bfY$ consists of a set of $n$ nodes and $m$ dyads $\{ Y_{ij}: i=1,\dots,n; j=1,\dots,n\}$ where 
$Y_{ij}=1$ if the pair $(i,j)$ is connected (full dyad), and $Y_{ij}=0$ otherwise (empty dyad). Edges connecting a node to itself 
are not allowed so $Y_{ii}=0$. The graph $\bfY$ may be directed (digraph) or undirected depending on the 
nature of the relationships between the nodes. 

Exponential random graph models (ERGMs) are a particular class of discrete linear exponential families which represent the probability distribution of $\bfY$ as
\begin{equation}
p(\bfy|\bftheta) 
   = \frac{q_{\bftheta}(\bfy)}
          {z(\bftheta)}
   = \frac{\exp\{\bftheta^T s(\bfy)\}}
          {\sum_{\bfy \in \mathcal{Y}} \exp\{\bftheta^T s(\bfy)\}}
\label{eq:ergm}
\end{equation}
where $s(\bfy)$ is a known vector of sufficient statistics computed on the network (or graph) (see \cite{sni:pat:rob:han06} and \cite{rob:pat:kal:lus07}) and $\bftheta$ are model parameters describing the dependence of $p(\bfy|\bftheta)$ on the observed statistics $s(\bfy)$. Estimating ERGM parameters is a challenging task due to the intractability of the normalising constant $z(\bftheta)$ and the issue of model degeneracy (see \cite{han03} and \cite{rin:fie:zho09}).

An important problem in many applications is the choice of the most appropriate set of explanatory network statistics $s(\bfy)$ to include in the model from a set of, \textit{a priori}, plausible ones. In fact in many applications there is a need to classify different types of networks based on the relevance of a set of configurations with respect to others.

From a Bayesian point of view, the model choice problem is transformed into one which aims to  estimate the posterior probability of all models within the considered class of competing models. In order to account for the uncertainty concerning the model selection process, Bayesian Model Averaging \citep{hoe:mad:raf:vol99} offers a coherent methodology which consists in averaging over many different competing models.

In the ERGM context, the intractability of the likelihood makes the use of standard techniques very challenging. The purpose of this paper is to present two new methods for Bayesian model selection for ERGMs.  This article is structured as follows. A brief overview of Bayesian model selection theory is given in Section \ref{sec:overview}. An across-model approach based on a trans-dimensional extension of the exchange algorithm of \cite{cai:fri11} is presented in Section \ref{sec:rjexchange}. The issue of the choosing parameters for the proposal distributions involved in the across model moves is addressed by presenting an automatic reversible jump exchange algorithm involving an independence sampler based on a distribution fitting a parametric density approximation to the within-model posterior. This algorithm bears some similarity to that presented in Chapter 6 of \cite{gre03}. We also present an approach to estimate the model evidence based on thermodynamic integration, although it is limited in that it can only be applied to ERGMs with a small number of parameters. This is outlined in Section~\ref{sec:evidence}.
% The second novel method is a within-model 
% approach for computing the model evidence. This approach is based on the path sampling approximation for estimating the likelihood 
% normalizing constant and it makes use of nonparametric density estimation procedures for approximating the posterior density of 
% each competing model (Section \ref{sec:evidence}). 
Three illustrations of how these new methods perform in practice are given in Section \ref{sec:apps}. Some conclusions are outlined in Section \ref{sec:conclusions}. The {\tt Bergm} package for {\tt R} \citep{cai:fri12b}, implements the newly developed methodology in this paper. It is available on the CRAN package repository at {\tt \href{http://cran.r-project.org/web/packages/Bergm/}{http://cran.r-project.org/web/packages/Bergm}}.

\section{Overview of Bayesian model selection}
\label{sec:overview}

Bayesian model comparison is commonly performed by estimating posterior model probabilities. More precisely, suppose that the competing models can be enumerated and indexed by the set $\{m_h : h = 1,\dots,H\}$. Suppose data $\bfy$ are assumed to have been generated by model $m_h$, the posterior distribution is:
\begin{equation}
p(\bftheta_h|\bfy,m_h) =
   \frac{p(\bfy|\bftheta_h,m_h) \; p(\bftheta_h|m_h)}
        {p(\bfy|m_h)},
\label{eq:posterior}
\end{equation}
where $p(\bfy|\bftheta_h,m_h)$ is the likelihood and $p(\bftheta_h|m_h)$ represents the prior distribution of the parameters of model $m_h$. The model evidence (or marginal likelihood) for model $m_h$, 
\begin{equation}
p(\bfy|m_h) = 
   \int_{\bftheta_h} p(\bfy|\bftheta_h,m_h) \; p(\bftheta_h|m_h) \; d\bftheta_h,
\label{eq:marginal}
\end{equation}
represents the probability of the data $\bfy$ given a certain model $m_h$ and is typically impossible to compute analytically. However, the model evidence is crucial for Bayesian model selection since it allows us to make statements about posterior model probabilities. Bayes' theorem can be written as 
\begin{equation}
p(m_h|\bfy) = 
  \frac{p(\bfy|m_h) \; p(m_h)}
       {\sum_1^H p(\bfy|m_h) \; p(m_h)}.
\end{equation}
Based on these posterior probabilities, pairwise comparison of models, $m_h$ and $m_k$ say, can be summarised by the posterior odds:
\begin{equation}
\frac{p(m_h|\bfy)}
     {p(m_k|\bfy)}
   =
   \frac{p(\bfy|m_h)}
        {p(\bfy|m_k)}
   \times
   \frac{p(m_h)}
        {p(m_k)}.
\label{eq:posteriorodds}
\end{equation}
This equation reveals how the data $\bfy$ through the Bayes factor 
\begin{equation}
BF_{hk} = 
   \frac{p(\bfy|m_h)}
        {p(\bfy|m_k)}
\label{eq:bayesfactor}
\end{equation}
updates the prior odds
\begin{equation}
   \frac{p(m_h)}
        {p(m_k)}
\end{equation}
to yield the posterior odds. Table~\ref{tab:kas_raf} displays guidelines which Kass and Raftery (\citeyear{kas:raf95}) suggest for interpreting Bayes factors.

\begin{table}[h]
\centering
\begin{tabular}{c|c}
\hline\hline
$BF_{hk}$ & Evidence against model $m_k$ \\
\hline
$1$ to $3$ & Not worth more than a bare mention \\
$3$ to $20$ & Positive \\
$20$ to $150$ & Strong \\
$>150$ & Very strong \\
\hline\hline
\end{tabular}
\caption{Guidelines for interpreting Bayes factors, following \cite{kas:raf95}.}
\label{tab:kas_raf}
\end{table}

By treating $p(m_h|\bfy)$ as a measure of the uncertainty of model $m_h$, a natural approach for model selection is to choose 
the most likely $m_h$, {\it a posteriori}, i.e. the model for which $p(m_h|\bfy)$ is the largest.

Bayesian model averaging \citep{hoe:mad:raf:vol99} provides a way of summarising model uncertainty in inference and prediction. 
After observing the data $\bfy$ one can predict a possible future outcome $\bfy^*$ by calculating an average of the posterior 
distributions under each of the models considered, weighted by their posterior model probability.:
\begin{equation}
p(\bfy^*|\bfy) = \sum_{h=1}^{H}p(\bfy^*|m_h,\bfy)p(m_h|\bfy),
\end{equation}
where $p(\bfy^*|m_h,\bfy)$ represents the posterior prediction of $\bfy^*$ according to model $m_h$ and data $\bfy$.

\subsection{Computing Bayes factors}

Generally speaking there are two approaches for computing Bayes factors: across-model and within-model estimation. The former strategy involves the use of an MCMC algorithm generating a single Markov chain which crosses the joint model and parameter space so as to sample from
\begin{equation}
p(\bftheta_h,m_h|\bfy)
\propto p(\bfy|\bftheta_h,m_h)\;p(\bftheta_h|m_h)\;p(m_h).
\label{eq:jointmodpar}
\end{equation}
One of the most popular approach used in this context is the reversible jump MCMC algorithm of \cite{gre95} which is briefly 
reviewed in Section \ref{sec:rjmcmc}. 
Within-model strategies focus on the posterior distribution (\ref{eq:posterior}) for each competing model $m_h$ separately, aiming to estimate their model evidence (\ref{eq:marginal}) which can then be used to calculate Bayes factors (see for example 
\cite{chi95}, \cite{chi:jel01}, \cite{nea01}, \cite{fri:pet08}, and \cite{fri:wys12}, who present a review of these methods). A within-model approach for estimating model evidence is presented in Section~\ref{sec:evidence}.

Across-model approaches have the advantage of avoiding the need for computing the evidence for each competing model by treating the model indicator $m_h$ as a parameter, but they require appropriate jumping design to produce computationally efficient and theoretically effective methods. Approximate Bayesian Computation (ABC) likelihood-free algorithms for model choice have been recently introduced by \cite{gre:rob:mar:rod:tal09} in order to allow the computation of the posterior probabilities of the models under competition. However these methods rely on proposing parameter values from the prior distributions which can differ very much from the posterior distribution and this can therefore affect the estimation process.
Variational approaches to Bayesian model selection have been presented by \cite{mcg:titt06} in the context of finite mixture distributions.

\subsubsection{Reversible jump MCMC}
\label{sec:rjmcmc}

The Reversible Jump MCMC (RJMCMC) algorithm is a flexible technique for model selection introduced by \cite{gre95} which allows simulation from target distributions on spaces of varying dimension. In the reversible jump algorithm, the Markov chain ``jumps'' between parameter subspaces (models) of differing dimensionality, thereby generating samples from the joint distribution of parameters and model indices. 

%RJMCMC can therefore be considered as a general framework which extends Metropolis-Hastings algorithms to variable dimension state spaces of the type $(\bftheta_l,m_l)$. In fact both simultaneous and single-site updating Metropolis-Hastings algorithms are special cases of RJMCMC.

To implement the algorithm we consider a countable collection of candidate models, $\{ m_k : k = 1, \dots, K \}$, each having an associated vector of parameters $\bftheta_k$ of dimension $D_k$ which typically varies across models. We would like to use MCMC to sample from the joint posterior  (\ref{eq:jointmodpar}).

In order to jump from $(\bftheta_k, m_k)$ to $(\bftheta_h, m_h)$, one may proceed by generating a random vector $\bfu$ from a 
distribution $g$ and setting $(\bftheta_h, m_h) = f_{kh}((\bftheta_k,m_k),\bfu)$. Similarly to jump from $(\bftheta_h, m_h)$ to 
$(\bftheta_k, m_k)$ we have $(\bftheta_k, m_k) = f_{hk}((\bftheta_h,m_h),\bfu^*)$ where $u^*$ is a random vector from a distribution 
$g^*$ and $f_{hk}$ is some deterministic function. However reversibility is only guaranteed when the parameter transition function 
$f_{kh}$ is a diffeomorphism, that is, both a bijection and its differential invertible. A necessary condition for this to apply is 
the so-called ``dimension matching'': $dim(\bftheta_k) + dim(\bfu) = dim(\bftheta_h) + dim(\bfu^*)$ (where $dim(\cdot)$ stands for 
``dimension of''). In this case the acceptance probability can be written as:
\begin{equation}
\min \left\lbrace 
   1, \frac{p(\bftheta_h,m_h|\bfy)}
           {p(\bftheta_k,m_k|\bfy)}
      \frac{p(m_h \rightarrow m_k)}
           {p(m_k \rightarrow m_h)}
      \frac{g^*(\bfu^*)}
           {g(\bfu)}
      |J|
     \right\rbrace 
\end{equation}
where $p(m_h \rightarrow m_k)$ is the probability of jumping from model $m_h$ to model $m_k$, and $|J|$ is the Jacobian resulting from the transformation from $((\bftheta_k,m_k),\bfu)$ to $((\bftheta_h,m_h),\bfu^*)$. 

Mixing is crucially affected by the choice of the parameters of the jump proposal distribution $g$ and this is one of the fundamental difficulties that makes RJMCMC often hard to use in practice \citep{bro:giu:rob03}.

\section{Reversible jump exchange algorithm}
\label{sec:rjexchange}

In the ERGM context, RJMCMC techniques cannot be used straightforwardly because the likelihood normalizing constant $z(\bftheta)$ in (\ref{eq:ergm}) cannot be computed analytically. 

Here we present an implementation of an RJMCMC approach for ERGMs based on an extension of the exchange algorithm of \cite{Murray06} developed for exponential random graph models.
The algorithm in \cite{cai:fri11} allows sampling within model $m_h$ from the following augmented distribution:
\begin{equation}
p(\bftheta'_h,\bfy',\bftheta_h|\bfy,m_h) \propto p(\bfy|\bftheta_h,m_h)p(\bftheta_h|m_h) h(\bftheta'_h|\bftheta_h,m_h) p(\bfy'|\bftheta'_h,m_h) 
\label{eq:exchange2}
\end{equation}
where $ p(\bfy|\bftheta_h,m_h)$ and $p(\bfy'|\bftheta'_h,m_h)$ are respectively the original likelihood defined on the observed data 
$\bfy$ and the augmented likelihood defined on simulated data $\bfy'$, $p(\bftheta_h|m_h)$ is the parameter prior and 
$h(\bftheta'_h|\bftheta_h,m_h)$ is any arbitrary proposal distribution for $\bftheta'_h$. Marginalising (\ref{eq:exchange2}) 
over $\bftheta'_h$ and $\bfy'$  yields the posterior of interest $p(\bftheta_h|\bfy,m_h)$.
Note that the simulation of a network $\bfy'$ from $p(\cdot|\bftheta'_h,m_h)$ is accomplished by a standard MCMC algorithm 
\citep{hun:han:but:goo:mor08} as perfect sampling has not yet been developed for ERGMs.

Auxiliary variable methods for intractable likelihood models, such as the exchange algorithm, have not been used in a trans-dimensional setting before. In order to propose to move from $(\bftheta_k,m_k)$ to $(\bftheta'_h,m'_h)$, the algorithm (\ref{eq:exchange2}) can be extended to sample from:
\begin{equation}
p(\bftheta'_h, \bftheta_k, m'_h, m_k, \bfy' | \bfy) 
   \propto 
   p(\bfy|\bftheta_k, m_k) p(\bftheta_k|m_k) p(m_k) h(\bftheta'_h,m'_h|\bftheta_k,m_k) p(\bfy'|\bftheta'_h, m'_h)
\label{eq:rjtarget}
\end{equation}
where $p(\bfy|\bftheta_k, m_k)$ and $p(\bfy'|\bftheta'_h, m'_h)$ are the two likelihood distributions for the data $\bfy$ under model $m_k$ and the auxiliary data $\bfy'$ under the competing model $m'_h$ respectively, $p(\bftheta_k|m_k)$ and $p(m_k)$ are the priors for the parameter $\bftheta_k$ and the respective model $m_k$ and $h(\bftheta'_h,m'_h|\bftheta_k,m_k)$ is some jump proposal distribution. 
Analogously as before, the marginal of (\ref{eq:rjtarget}) for $\bftheta'_h$ and $m'_h$ is the distribution of interest (\ref{eq:jointmodpar}).

Suppose that the current state of the chain is $(\bftheta_k,m_k)$ and let us propose a move to $(\bftheta'_h,m'_h)$. The Metropolis-Hastings ratio for accepting the whole move is:
\begin{equation*}
\begin{split}
&
   \frac{p(\bfy'|\bftheta_k, m_k)}
        {p(\bfy|\bftheta_k, m_k)}
   \frac{p(\bfy|\bftheta'_h, m'_h)}
        {p(\bfy'|\bftheta'_h, m'_h)}
   \frac{p(\bftheta'_h|m'_h)}
        {p(\bftheta_k|m_k)}
   \frac{p(m'_h)}
        {p(m_k)}
   \frac{h(\bftheta_k,m_k|\bftheta'_h,m'_h)}
        {h(\bftheta'_h,m'_h|\bftheta_k,m_k)}
=\\
&
   \frac{q_{\bftheta_k,m_k}(\bfy')}
        {q_{\bftheta_k,m_k}(\bfy)}
   \frac{q_{\bftheta'_h,m'_h}(\bfy)}
        {q_{\bftheta'_h,m'_h}(\bfy')}
   \frac{p(\bftheta'_h|m'_h)}
        {p(\bftheta_k|m_k)}
   \frac{p(m'_h)}
        {p(m_k)}
   \frac{h(\bftheta_k,m_k|\bftheta'_h,m'_h)}
        {h(\bftheta'_h,m'_h|\bftheta_k,m_k)}
   \times
   \frac{z(\bftheta_k)}
        {z(\bftheta_k)}
   \frac{z(\bftheta'_h)}
        {z(\bftheta'_h)}
\end{split}
\end{equation*}
where $q_{\bftheta_k,m_k}(\bfy)$ indicates the unnormalised likelihood of $p(\bfy|\bftheta_k,m_k)$ (and so forth for the other 
functions $q(\cdot)$). Note that the normalising constants corresponding to the unnormalised likehoods cancel. Therefore the ratio 
above is free of any dependence on normalising constants and so can be evaluated.

The issue with this method is that tuning the jump proposals $h(\cdot)$ in a sensible way so as to get a reasonable mixing can be difficult and automatic choice of jump parameters \citep{bro:giu:rob03} does not apply in this context due to the intractability of the likelihood distribution.

\subsection{Pilot-tuned RJ exchange algorithm}
\label{sec:pilot-tuned}

We now consider nested models or models differing by at most one variable. In this case, the move from $(\bftheta_k,m_k)$ to a 
larger model $(\bftheta'_{k+1},m'_{k+1})$ such that $dim(m'_{k+1}) = dim(m_k)+1$ can be done by proposing the transformation 
$(\bftheta'_{k+1},m'_{k+1}) = ((\bftheta_k,\theta'_{k+1}),m_{k+1})$ where the $(k+1)$-th parameter value $\theta'_{k+1}$ is 
generated from some distribution $g_{k+1}$ and then accepting the move with the following probability:
\begin{equation*}
\alpha=
\min \left\lbrace 
1,
   \frac{q_{\bftheta_k,m_k}(\bfy')}
        {q_{\bftheta_k,m_k}(\bfy)}
   \frac{q_{\bftheta'_{k+1},m'_{k+1}}(\bfy)}
        {q_{\bftheta'_{k+1},m'_{k+1}}(\bfy')}
   \frac{p(\bftheta'_{k+1}|m'_{k+1})}
        {p(\bftheta_k|m_k)}
   \frac{p(m'_{k+1})}
        {p(m_k)}
   \frac{1}
        {g_{k+1}(\theta'_{k+1})}
   \frac{h(m_k|m'_{k+1})}{h(m'_{k+1}|m_k)}
     \right\rbrace.
\label{eq:pilott}
\end{equation*}
The reverse move is accepted with a probability based upon the reciprocal of the acceptance ratio (\ref{eq:pilott}). The jump within the same model $m_k$ is accepted with the following probability:
\begin{equation*}
\alpha=
\min \left\lbrace 
1,
   \frac{q_{\bftheta_k,m_k}(\bfy')}
        {q_{\bftheta_k,m_k}(\bfy)}
   \frac{q_{\bftheta'_{k},m'_{k}}(\bfy)}
        {q_{\bftheta'_{k},m'_{k}}(\bfy')}
   \frac{p(\bftheta'_k|m'_k)}
        {p(\bftheta_k|m_k)}
   \frac{p(m'_k)}
        {p(m_k)}
   \frac{g(\bftheta_k)}
        {g(\bftheta'_k)}
     \right\rbrace.
\end{equation*}

\subsection{Auto-RJ exchange algorithm}

Finding suitable parameter values for the proposals for the jump move between models is a very challenging task and is vital in order to ensure
adequate mixing of the trans-dimensional Markov chain. In practice, tuning the parameters of the proposals for the trans-dimensional move is very difficult without any information about the posterior density covariance structure. 
In our experience, in the context of ERGMs, it is extremely difficult to pilot tune a RJMCMC approach to yield adequate mixing rates, rendering this approach impractical for most situations.
A possible approach would be to 
use an independence sampler which does not depend on the current state of the MCMC chain but fits a parametric density 
approximation to the within-model posterior distribution so as to have an acceptance rate as high as possible.

In this spirit, we can propose to jump from $(\bftheta_k,m_k)$ to $(\bftheta'_h,m'_h)$ using the following jump proposals:
\begin{equation}
h(\bftheta'_h,m'_h|\bftheta_k,m_k) = w(\bftheta'_h|m'_h)\;h(m'_h|m_k)
\label{eq:indjumpprop}
\end{equation}
where $h(m_k|m'_h)$ represents a between-model jump proposal from model $m_k$ to model $m'_h$ and $w(\bftheta'_h|m'_h)$ is the 
within-model jump proposal for model $m'_h$. As remarked above, the within-model proposals require careful tuning. Posterior density 
approximations such as standard distributions with parameters determined by the moments of a sample drawn from (\ref{eq:rjtarget}) 
can be used as within model proposals for each competing model. Indeed this is similar to the type of strategy outlined in Chapter 6 of 
\cite{gre03}. For example, $w(\bftheta_l|m_l)$ can be a normal distribution $\mathcal{N}(\hat{\bfmu}_{l},\hat{\bfSigma}_{l})$ where 
$\hat{\bfmu}_{l}$ and $\hat{\bfSigma}_{l}$ are the posterior mean and covariance estimates for each model $m_l$. In our experience 
the choice of normal proposals appear to fit quite well in most of the examples we looked at, although using $t-$distributions
may be more robust to heavier tails in the posterior. 

The algorithm can be therefore summarized in two steps: the first step (offline) is used to sample from the posterior 
(\ref{eq:exchange2}) of each model $m_l$ and to estimate the parameters $\hat{\bfmu}_{l}$ and $\hat{\bfSigma}_{l}$ of the within-model 
jump proposal; the second step (online) carries out the MCMC computation of (\ref{eq:rjtarget}).

The algorithm can be written in the following concise way:
\begin{algorithm}
\step{OFFLINE RUN}
\step{(0)} Estimation of $p(\bftheta_l| \bfy, m_l) \;$ for $l=1,\dots,H$\\
\textit{i} Set $\hat{\bfmu}_{l} = \EE(\bftheta_l|\bfy, m_l)$ and $\hat{\bfSigma}_{l} = Cov(\bftheta_l|\bfy, m_l)$ \\
\textit{ii} Use $w(\bftheta_l|m_l) \sim \mathcal{N}(\hat{\bfmu}_{l},\hat{\bfSigma}_{l})$ as within-model jump proposals, when proposing to jump to model $m_l$\\
\step{ONLINE RUN}
\step{(1.1)} Gibbs update of $(m'_h,\bftheta'_h, \bfy')$\\
\textit{i} Propose $m'_h$ from the prior $p(\cdot)$\\
\textit{ii} Propose $\bftheta'_h$ with probability $w(\cdot|\hat{\bfmu}_{h},\hat{\bfSigma}_{h})$\\
\textit{iii} Draw $\bfy'$ from $p(\cdot|\bftheta'_h,m'_h)$
\step{(1.2)}Accept the jump from $(\bftheta_k, m_k)$ to $(\bftheta'_h, m'_h)$ with probability:
\begin{equation*}
\min \left\lbrace 
1,
   \frac{q_{\bftheta_k,m_k}(\bfy')}
        {q_{\bftheta_k,m_k}(\bfy)}
   \frac{q_{\bftheta'_h,m'_h}(\bfy)}
        {q_{\bftheta'_h,m'_h}(\bfy')}
   \frac{p(\bftheta'_h|m'_h)}
        {p(\bftheta_k|m_k)}
   \frac{p(m'_h)}
        {p(m_k)}
   \frac{w(\bftheta_k|\hat{\bfmu}_{k},\hat{\bfSigma}_{k})}
        {w(\bftheta'_h|\hat{\bfmu}_{h},\hat{\bfSigma}_{h})}
   \frac{h(m_k|m'_k)}{h(m'_h|m_k)}
     \right\rbrace. 
\end{equation*}
\end{algorithm}

\section{Estimating model evidence}
\label{sec:evidence}

In this section we present a  within-model approach for estimating the evidence $p(\bfy)$ (For ease of notation, we will 
omit the conditioning on the model indicator $m_l$). The aim is to provide a useful method for low-dimensional models to 
use as a ``ground-truth'' reference to compare with the reversible jump exchange algorithm. The method follows from noticing 
that for any parameter $\bftheta^\star$, equation (\ref{eq:posterior}) implies that:
\begin{equation}
p(\bfy) 
=
   p(\bfy|\bftheta^\star) 
   \frac{p(\bftheta^\star)}
        {p(\bftheta^\star|\bfy)}
=
   \frac{q_{\bftheta^\star}(\bfy)}
        {z(\bftheta^\star)}
   \frac{p(\bftheta^\star)}
        {p(\bftheta^\star|\bfy)}.
\label{eq:evidence}
\end{equation}
This is also the starting point for Chib's method for estimating the evidence \citep{chi95}.
Typically $\bftheta^\star$ is chosen as a point falling in the high posterior probability region so as to increase the 
accuracy of the estimate. To estimate (\ref{eq:evidence}), the calculation of the intractable likelihood normalizing 
constant $z(\bftheta^\star)$ and an estimate of the posterior density $p(\bftheta^\star |\bfy)$ are required.

\subsubsection*{Estimating $z(\bftheta^\star)$ via path sampling}

The first problem can be tackled using a path sampling approach \citep{gel:men98}. Consider introducing an auxiliary variable $t \in [0,1]$. We consider the following distribution:
\begin{equation}
p_t(\bfy|\bftheta) =
p(\bfy|\bftheta)^t = 
p(\bfy|\bftheta t) 
   = \frac{q_{\bftheta t}(\bfy)}
          {z(\bftheta t)}
   = \frac{\exp\{(\bftheta t)^T s(\bfy)\}}
          {\sum_{\bfy \in \mathcal{Y}} \exp\{(\bftheta t)^T s(\bfy)\}}.
\end{equation}
Taking logarithms and differentiating $\log \left[ z(\bftheta^\star t) \right]$ with respect to $t$ yields:
\begin{align}
\frac{d}
     {dt}
\log\left[ z(\bftheta^\star t)\right] \notag
&=
\frac{1}
     {z(\bftheta^\star t)}
\frac{d}
     {dt}
z(\bftheta^\star t)\\ \notag
&= 
\frac{1}
     {z(\bftheta^\star t)}
\frac{d}
     {dt}
\sum_{\bfy \in \mathcal{Y}} 
\exp\left\lbrace (\bftheta^{\star} t)^T s(\bfy)\right\rbrace\\ \notag
&=
\frac{1}
     {z(\bftheta^\star t)}
\sum_{\bfy \in \mathcal{Y}} 
\left[ \bftheta^{\star T} s(\bfy) \right]
\exp\left\lbrace(\bftheta^{\star} t)^T s(\bfy)\right\rbrace\\ \notag
&=
\sum_{\bfy \in \mathcal{Y}} 
\left[ \bftheta^{\star T} s(\bfy) \right]\;p(\bfy|\bftheta^\star t)\\
&=\EE_{\bfy|\bftheta^{\star} t}\left[\bftheta^{\star T} s(\bfy)\right].
\label{eq:pathsampling}
\end{align}
where $\EE_{\bfy|\bftheta^{\star} t}$ denotes the expectation with respect to the sampling distribution $p(\bfy|\bftheta^{\star} t)$. 
Therefore integrating (\ref{eq:pathsampling}) from $0$ to $1$ gives:
\begin{equation*}
\log\left\lbrace\frac{z(\bftheta^\star)}{z(\bf0)}\right\rbrace \notag
   = \int\limits_{0}^{1} \EE_{\bfy|\bftheta^{\star} t}[\bftheta^{\star T}s(\bfy)]\; dt.
\label{eq:log1star}
\end{equation*}
Now if we choose a discretisation of the variable $t$ such that $t_0 = 0 < \dots < t_i < \dots < t_I = 1$, this leads to the 
following approximation:
\begin{equation}
\log\left\lbrace\frac{z(\bftheta^\star)}{z(\bf0)}\right\rbrace
   \approx \sum\limits_{i=0}^{I-1} (t_{i+1} - t_{i})\; 
   \left(
   \frac{\EE_{\bfy|\bftheta^{\star} t_i}[\bftheta^{\star T}s(\bfy)] + \EE_{\bfy|\bftheta^{\star} t_{i+1}}[ \bftheta^{\star T} 
s(\bfy)]}{2} \right).
\label{eq:logzratio}
\end{equation}
Remember that $z(\mathbf{0})$ is analytically available 
and it is equal to $2^{\binom{n}{2}}$ i.e. the number of possible graphs on the $n$ nodes of the observed network. In terms 
of computation, $\EE_{\bfy|\bftheta^\star t_i}[\bftheta^{\star T}s(\bfy)]$ can be easily estimated using the same procedures 
used for simulating auxiliary data from the ERGM likelihood. Hence in (\ref{eq:logzratio}) two types of error emerge: 
discretisation of (\ref{eq:evidence}) and Monte Carlo error due to the simulation approximation of $\EE_{\bfy|\bftheta^\star t_i}[\bftheta^{\star T}s(\bfy)]$.
The path of $t_i$'s is important for the efficiency of the evidence estimate. For example, we can choose a path of the type 
$t_i = (1/I)^c$ where $c$ is some tuning constant: for $c=1$ we have equal spacing of the $I$ points in the interval $[0,1]$, 
for $c > 1$ we have that the $t_i$'s are chosen with high frequency close to $0$ and for $0 < c < 1$ we have that the $t_i$'s 
are chosen with high frequency close to $1$.

\subsubsection*{Estimating $p(\bftheta^\star|\bfy)$}

A sample from the posterior $p(\bftheta | \bfy)$ can be gathered (via the exchange algorithm, for example) and used to calculate 
a kernel density estimate of the posterior probability at the point $\bftheta^{\star}$. In practice, because of the curse of dimensionality,  
this implies that the method cannot be used, for models with greater than $5$ parameters.
In this paper we used the fast and easy to use {\tt np} package for {\tt R} \citep{hay:rac08} to perform a nonparametric density 
estimation of the posterior $p(\bftheta^\star|\bfy)$.

\begin{figure}[htp]
\centering
\includegraphics[scale=0.8]{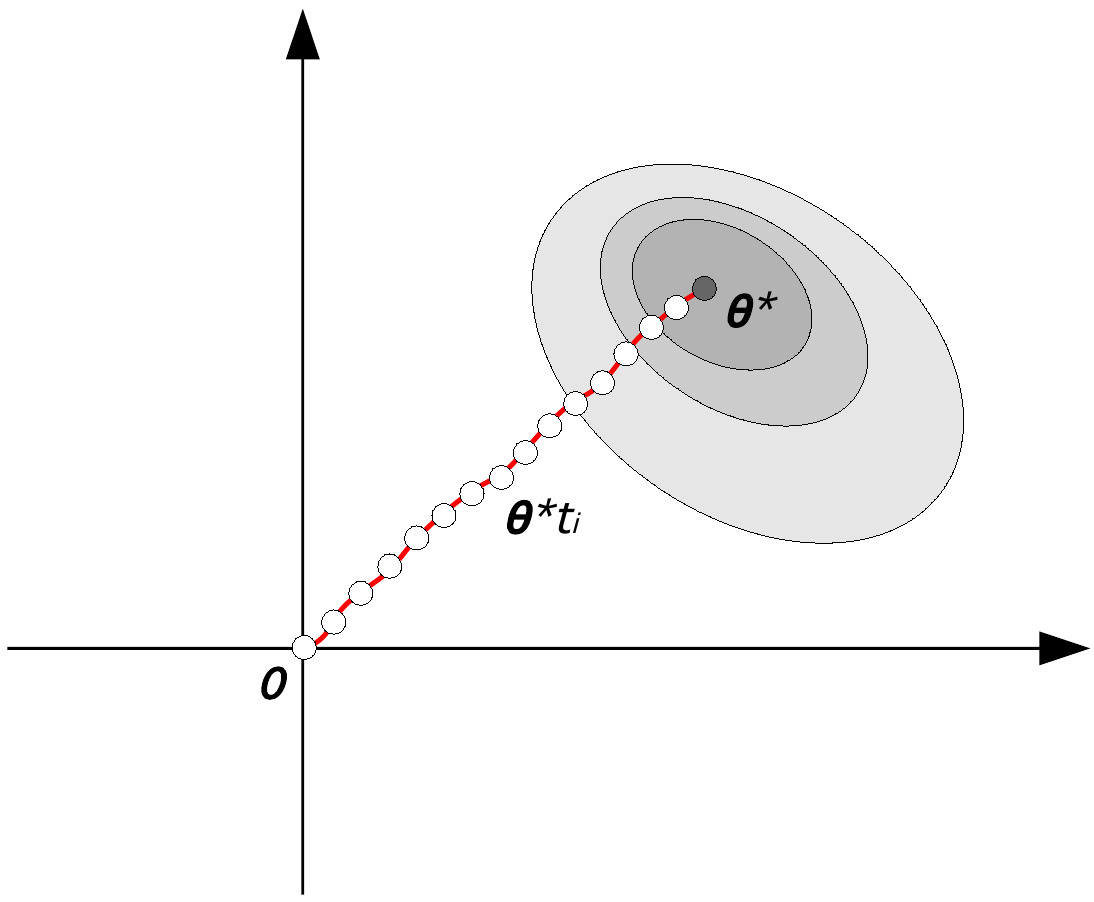}
\caption{Path sampling: for each $\bftheta^\star$ we estimate $z(\bftheta^\star)$ via path sampling using the expected network 
statistics simulated from some points $\bftheta^\star t_i$ along the line connecting $\bf0$ to $\bftheta^\star$.}
\label{fig:path}
\end{figure}

\section{Applications}
\label{sec:apps}

\subsection{Gahuku-Gama system}

\begin{figure}[htp]
\centering
\fbox{
\includegraphics[scale=.6]{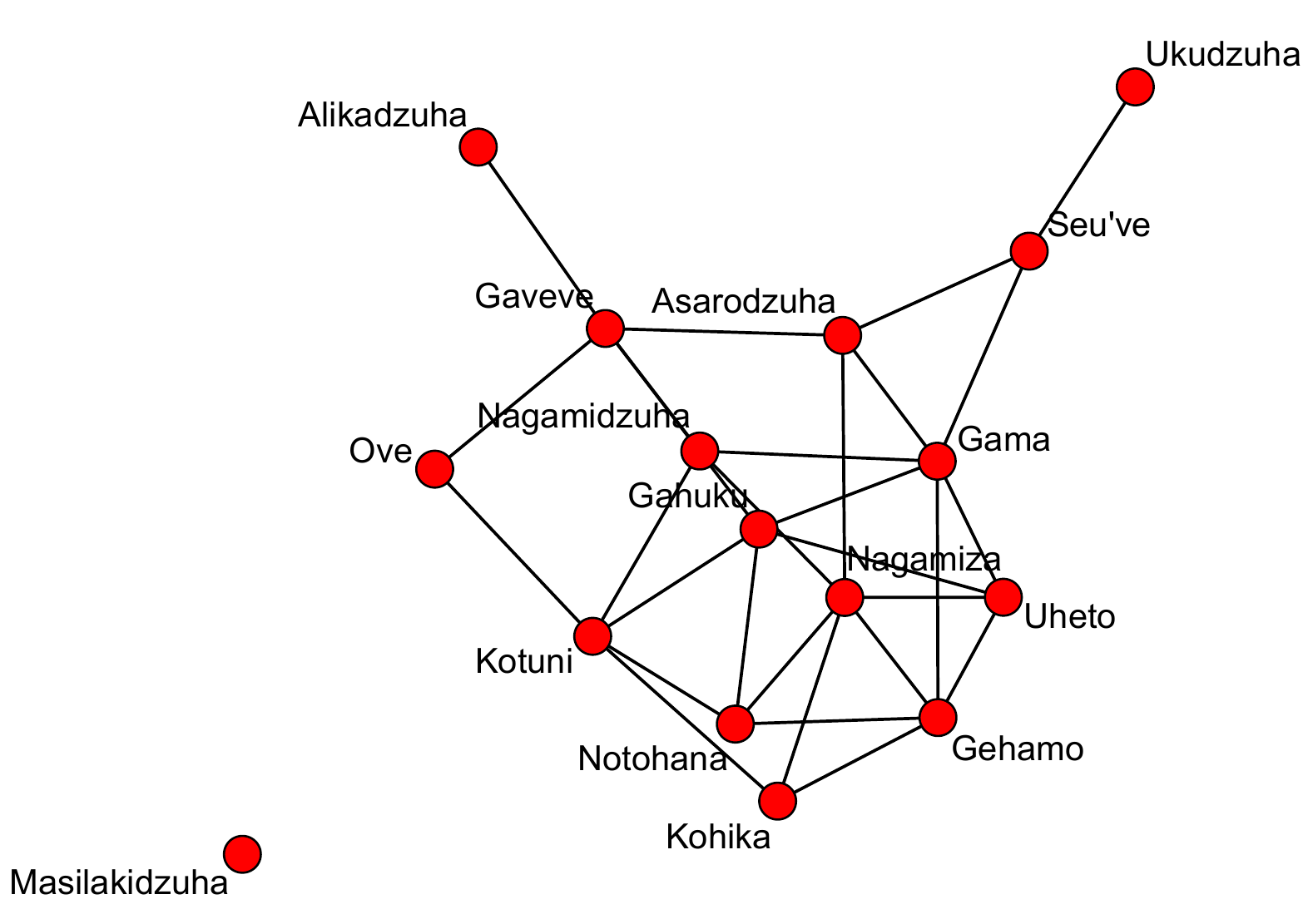}
}\\\vspace{.5cm}
\fbox{
\includegraphics[scale=.6]{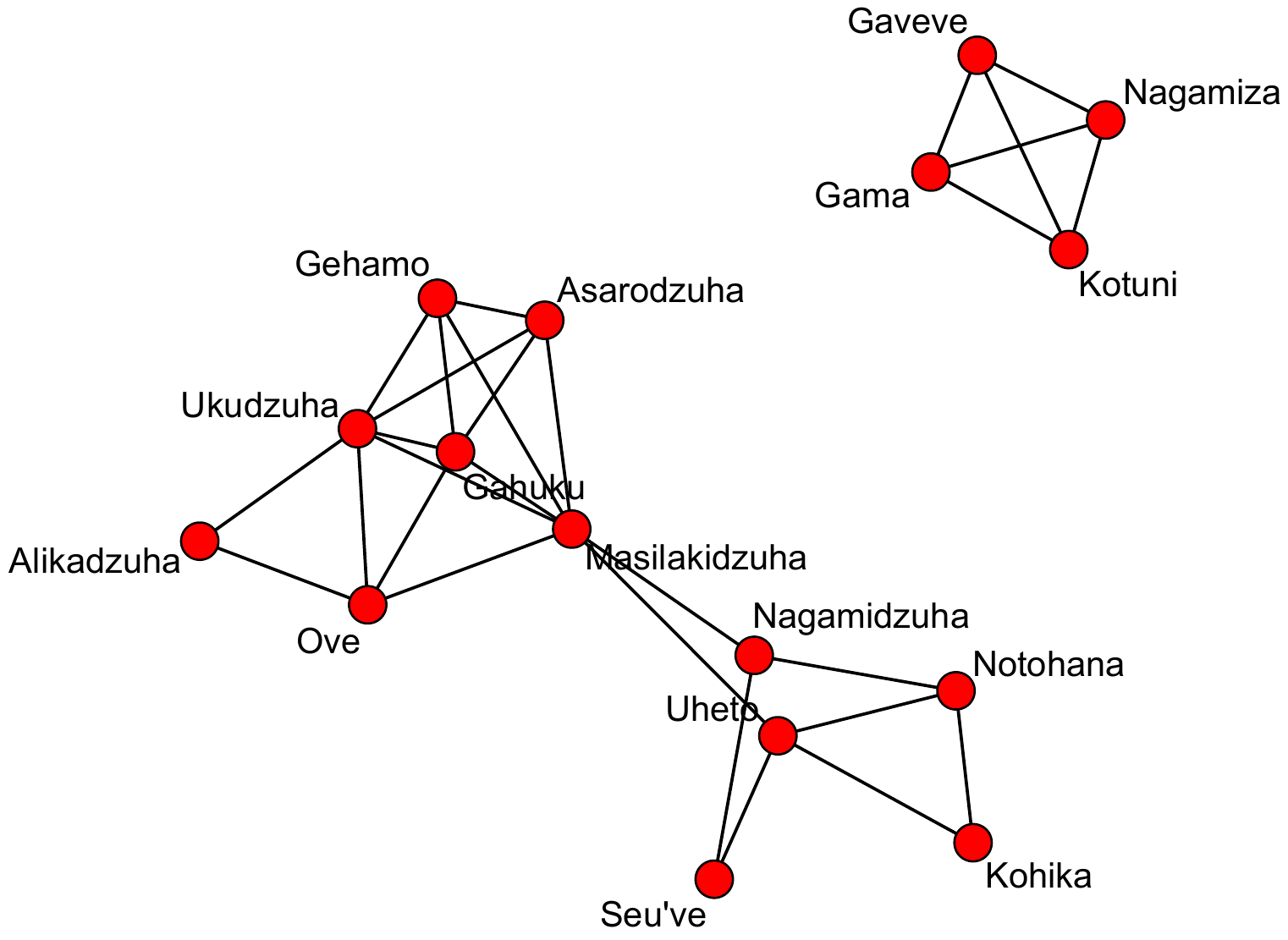}
}
\caption{Gahuku-Gama system graphs: Gamaneg (top) and Gamapos (bottom).}
\label{fig:gama_graph}
\end{figure}

The Gahuku-Gama system \citep{rea54} of the Eastern Central Highlands of New Guinea was used by \cite{hag:har84} to describe an 
alliance structure among 16 sub-tribes of Eastern Central Highlands of New Guinea (Figure \ref{fig:gama_graph}). The system has 
been split into two network: the ``Gamaneg'' graph for antagonistic (``hina") relations and the ``Gamapos'' for alliance (``rova") 
relations. An important feature of these structures is the fact that the enemy of an enemy can be either a friend or an enemy.

\subsubsection{Gamaneg}

We first focus on the Gamaneg network by using the 3 competing models specified in Table \ref{tab:gamaneg_modprop} using the following network statistics:
\begin{center}
\begin{tabular}{ll}
edges & $\sum_{i<j}y_{ij}$\\
triangles & $\sum_{i<j<k}y_{jk}y_{ik}y_{ij}$\\ 
4-cycle & $\sum_{i<j<l<k}y_{ij}y_{jl}y_{lk}y_{ki}$
\end{tabular}
\end{center}
We are interested to understand if the transitivity effect expressed by triad closure (triangle) and 4-cycle which is a closed structure that permits to measure the dependence between two edges that do not share a node \citep{pat:rob02}.

\begin{table}[htp]
\centering
\begin{tabular}{ll}
\hline\hline
Model $m_1$ & $\bfy \sim$ edges\\
Model $m_2$ & $\bfy \sim$ edges $+$ triangles\\
Model $m_3$ & $\bfy \sim$ edges $+$ triangles $+$ 4-cycle\\
\hline\hline
\end{tabular}
\caption{Competing models.}
\label{tab:gamaneg_modprop}
\end{table}

\begin{figure}[htp]
\centering
\includegraphics[scale=0.5]{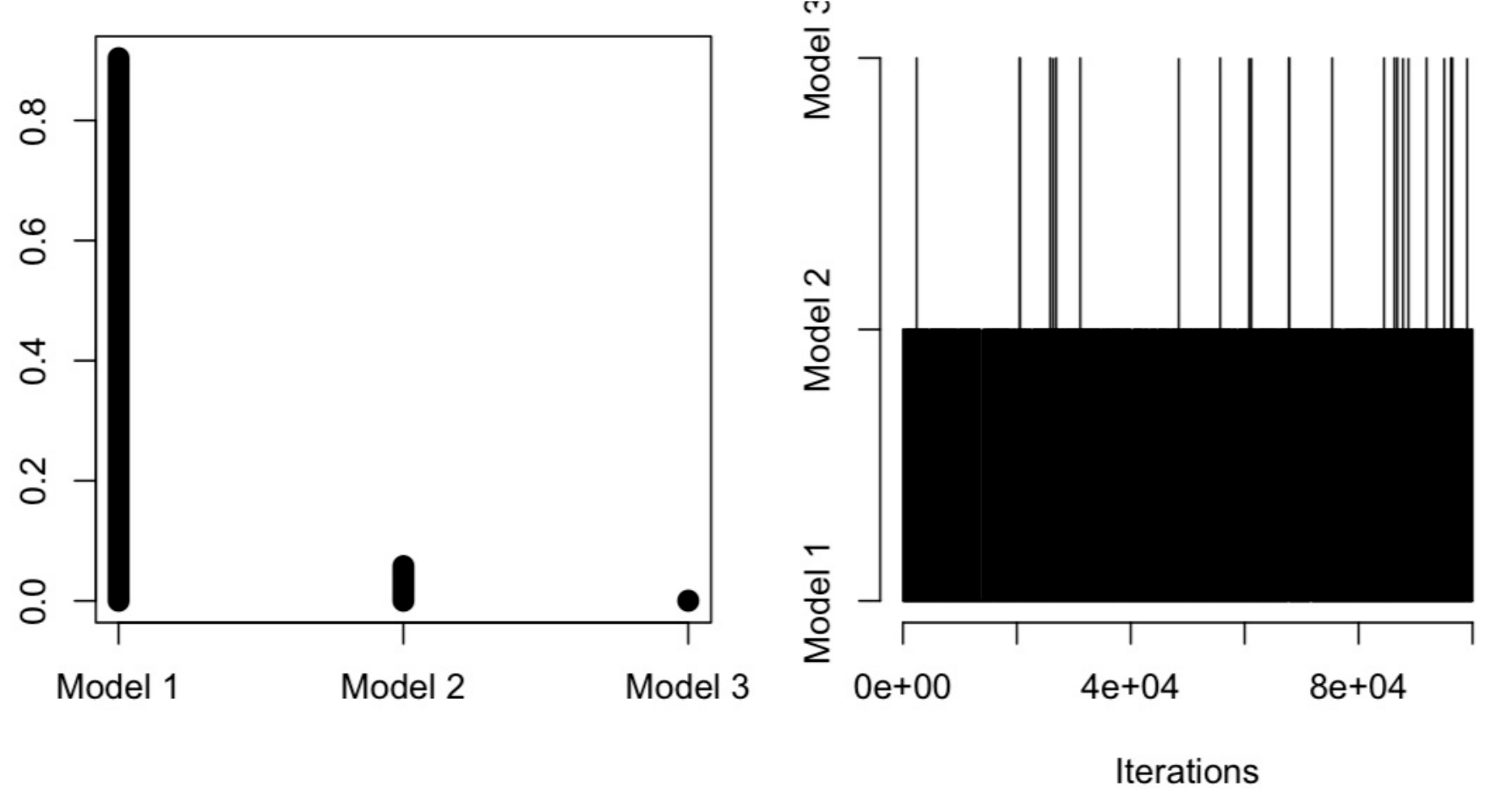}\\
\vspace{1.5cm}
\includegraphics[scale=0.8]{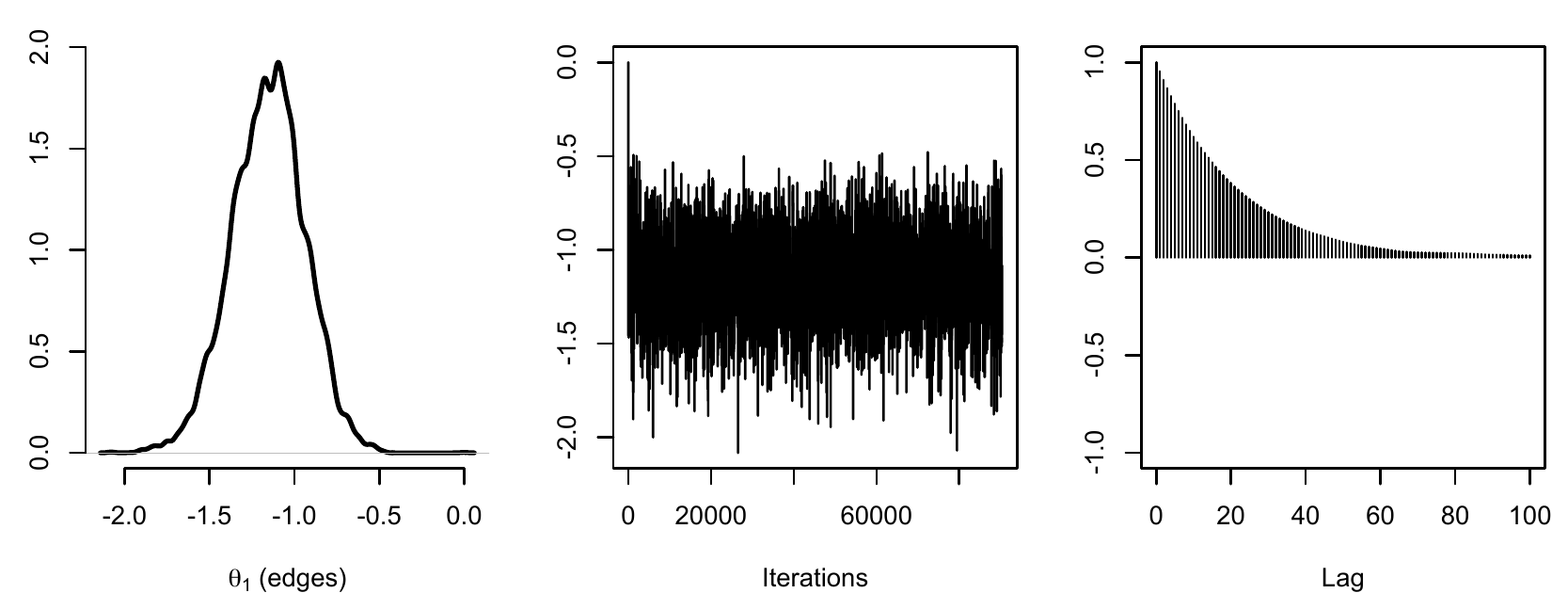}
\caption{Pilot-tuned RJ exchange algorithm output: posterior model probabilities (top) and posterior parameter probabilities for model $m_1$ (bottom).}
\label{fig:gamaneg_post_pt}
\end{figure}

\begin{figure}[htp]
\centering
\includegraphics[scale=0.5]{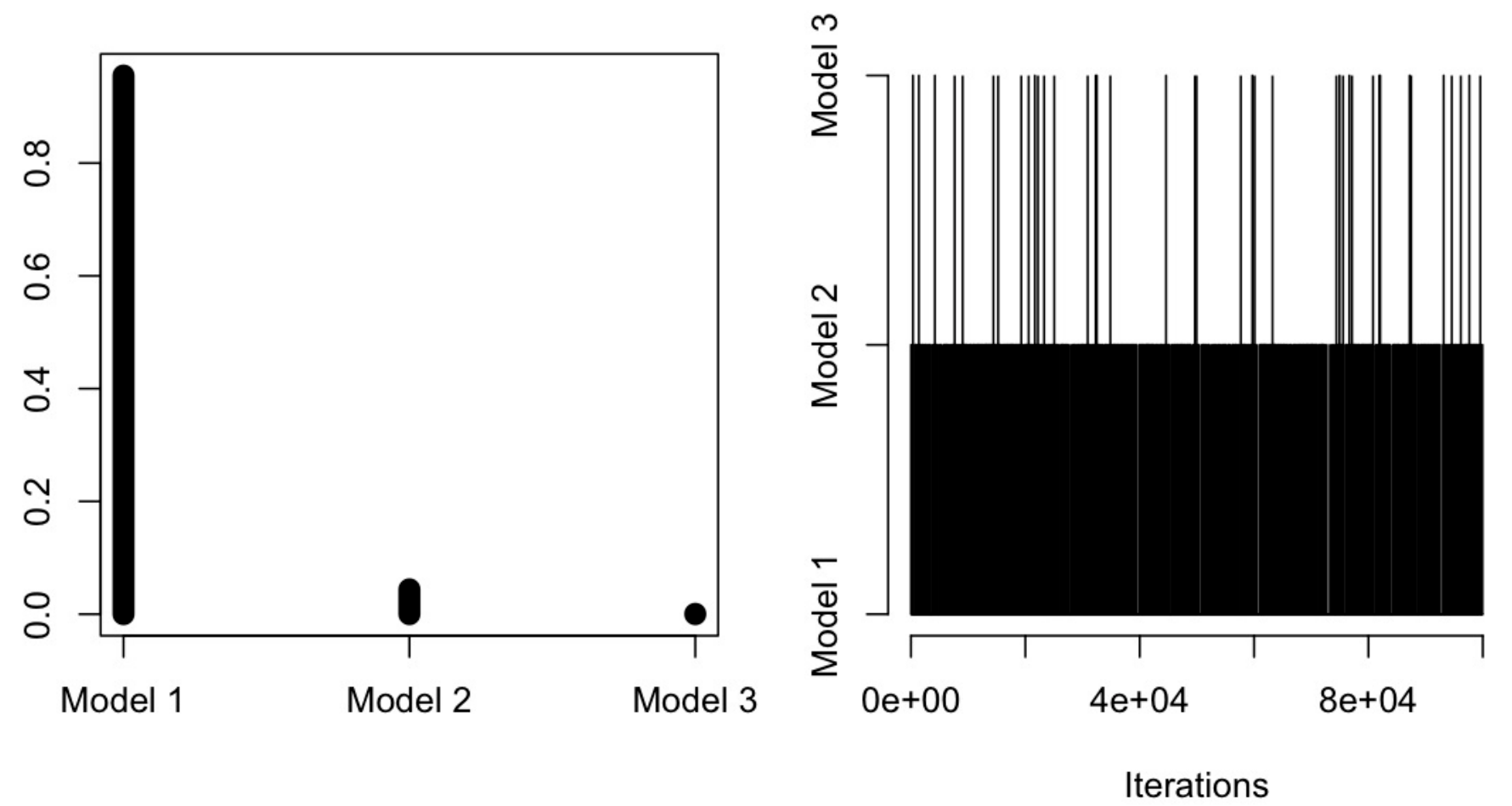}\\
\vspace{1cm}
\includegraphics[scale=0.8]{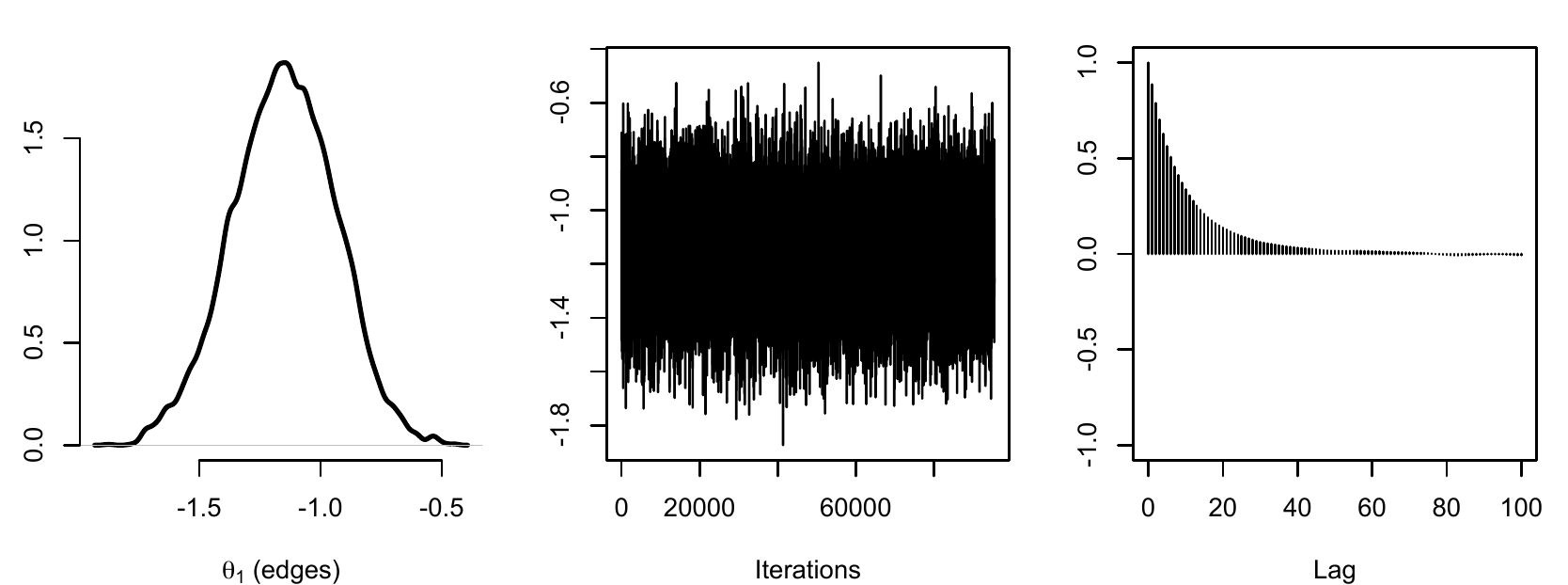}
\caption{Auto-RJ exchange algorithm output: posterior model probabilities (top) and posterior parameter probabilities for model $m_1$ (bottom).}
\label{fig:gamaneg_post_auto}
\end{figure}

Both the pilot-tuned RJ and auto-RJ exchange algorithms were run for  $100,000$ iterations using very flat normal parameter priors 
$p(\bftheta_l|m_l) \sim \mathcal{N}(0,100\bfI_l)$ for each model $m_l$ where $\bfI_l$ is the identity matrix of size equal to the 
number of dimensions of model $m_l$ and $3,000$ iterations for the auxiliary network simulation. The proposal distributions 
of the pilot-tuned RJ were empirically tuned so as to get reasonable acceptance rates for each competing model. The offline 
step of the auto-RJ consisted of gathering an approximate sample from $p(\bftheta|\bfy)$ and then estimating the posterior 
moments $\hat{\bfmu}_{l}$ and $\hat{\bfSigma}_{l}$ for each of the three models. The exchange algorithm was run for 
$1,000 \times D_{l}$ iterations (discarding the first $100 \times D_{l}$ iterations as burn-in) where $D_{l}$ is the dimension of 
the $l$-th model using the population MCMC approach described in \cite{cai:fri11}. The accuracy of the estimates $\hat{\bfmu}_{l}$ 
and $\hat{\bfSigma}_{l}$ depends on the number of iterations of the auto-RJ offline run. In this example, the above number of 
iterations $1,000 \times D_{l}$ of has been empirically shown to be sufficient for each competing model $m_l$. 
In this example and all the examples that follow we use uniform model prior and uniform between-model jump proposals. 
Tables \ref{tab:gamaneg_parpost} and \ref{tab:gamaneg_rates} report the posterior parameter estimates of the model selected for 
the pilot-tuned RJ and auto-RJ. 
From these tables we can see that the pilot-tuned RJ sampler exhibits poor within-model mixing with respect to the good mixing of 
the auto-RJ sampler. This greatly affected the convergence of the pilot-tuned RJ leading to very poor posterior estimates. 
Figure \ref{fig:gamaneg_post_pt} shows the results 
from the pilot-tuned RJ, namely, model posterior diagnostic plots and the parameter posterior diagnostic plots. 
Figure \ref{fig:gamaneg_post_auto} shows the same plots from auto-RJ. Between-model and within-model acceptance rates 
(reported in Table \ref{tab:gamaneg_rates}) are calculated as the proportions of accepted moves from $(\bftheta_k,m_k)$ to 
model $(\bftheta'_h,m'_h)$ for each $k : k \neq h$ and when $k = h$, respectively. The mixing of the auto-RJ algorithm within each model 
is faster than the pilot-tuned RJ algorithm due to the good approximation to the posterior distribution. The pilot-tuned 
algorithm took about 24 minutes to complete the estimation and the auto-RJ took about 31 minutes (including the offline step).

\begin{table}[htp]
\centering
\begin{tabular}{l|cc|cc}
\hline\hline
    & \multicolumn{2}{c|}{Pilot-tuned RJ} & \multicolumn{2}{c}{Auto-RJ}\\
Parameter & Post. Mean & Post. Sd. & Post. Mean & Post. Sd.\\
\hline
\multicolumn{5}{c}{Model $m_1$}\\
\hline
$\theta_1$ (edge)      & -1.15 & 0.21 & -1.15 & 0.21\\
\hline
\multicolumn{5}{c}{Model $m_2$}\\
\hline
$\theta_1$ (edge)      & -0.97 & 0.36 & -0.96 & 0.37\\
$\theta_2$ (triangle)  & -0.31 & 0.41 & -0.29 & 0.37\\
\hline
\multicolumn{5}{c}{Model $m_3$}\\
\hline
$\theta_1$ (edge)      & -0.98 & 0.51 & -1.15 & 0.37\\
$\theta_2$ (triangle)  & -0.76 & 0.47 & -0.31 & 0.42\\
$\theta_3$ (4-cycle)   & -0.05 & 0.12 &  0.02 & 0.17\\
\hline\hline
\end{tabular}
\caption{Summary of posterior parameter estimates.}
\label{tab:gamaneg_parpost}
\end{table}

\begin{table}[htp]
\centering
\begin{tabular}{l|c|c}
\hline\hline
Within-model & Pilot-tuned RJ & Auto-RJ\\
\hline
Model $m_1$ & $0.14$ & $0.62$\\
Model $m_2$ & $0.11$ & $0.42$\\
Model $m_3$ & $0.00$  & $0.48$\\
\hline\hline
Between-model & $0.07$ & $0.04$\\
\hline\hline
\end{tabular}
\caption{Acceptance rates.}
\label{tab:gamaneg_rates}
\end{table}

\begin{table}[htp]
\centering
\begin{tabular}{l|c|c}
\hline\hline
           & Pilot-tuned RJ & Auto-RJ\\
\hline
$\widehat{BF}_{1,2}$ & $14.46$   & $21.68$ \\
$\widehat{BF}_{1,3}$ & $1506.43$ & $1425.77$\\
\hline\hline
$\widehat{p}(m_1|\bfy)$ & $0.93$ & $0.95$\\
$\widehat{p}(m_2|\bfy)$ & $0.06$ & $0.04$\\
$\widehat{p}(m_3|\bfy)$ & $0.01$ & $0.01$\\
\hline\hline
\end{tabular}
\caption{Bayes factor and posterior model probability estimates.}
\label{tab:gamaneg_bf}
\end{table}

\begin{figure}[htp]
\centering
\includegraphics[scale=0.4]{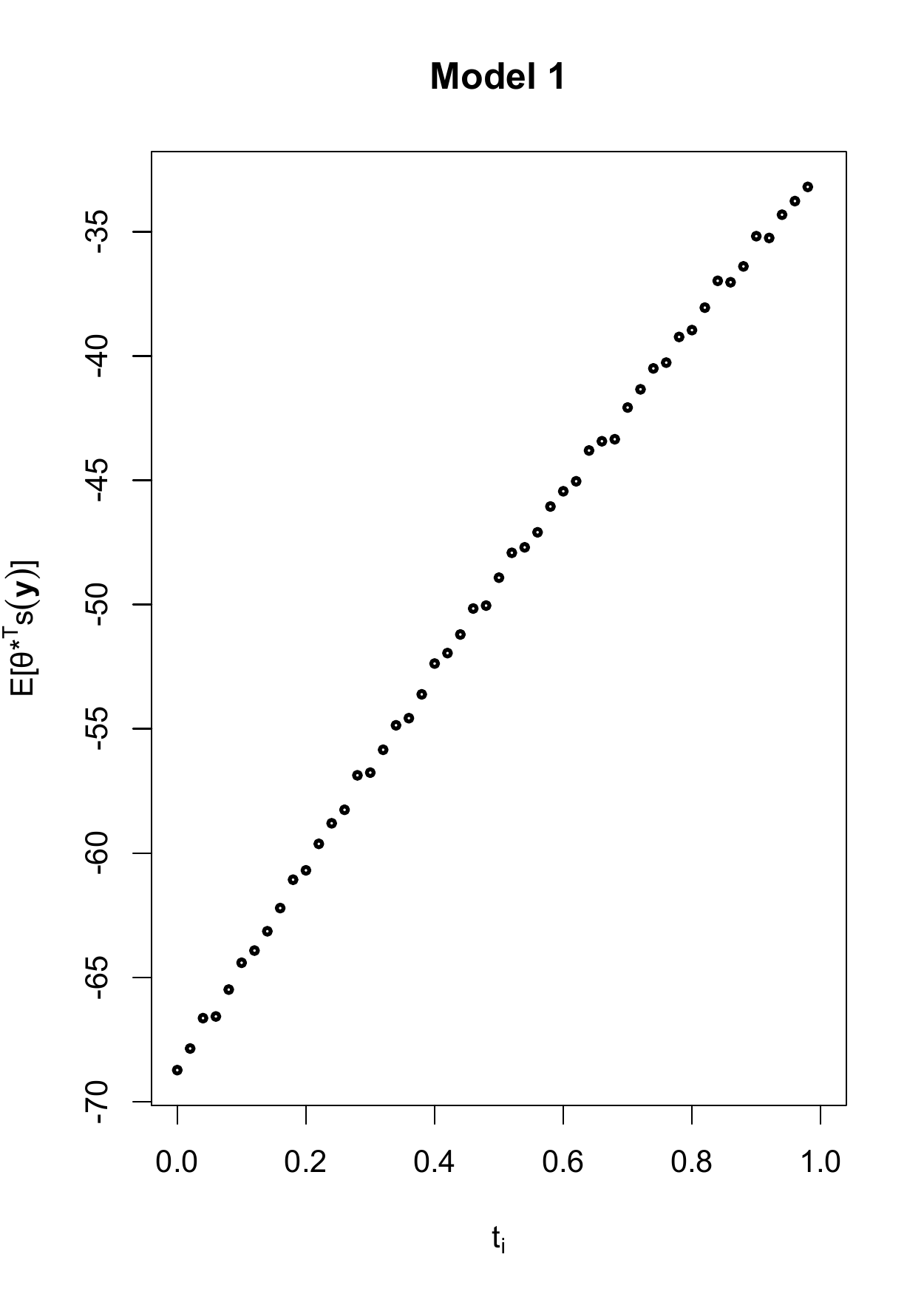}\quad
\includegraphics[scale=0.4]{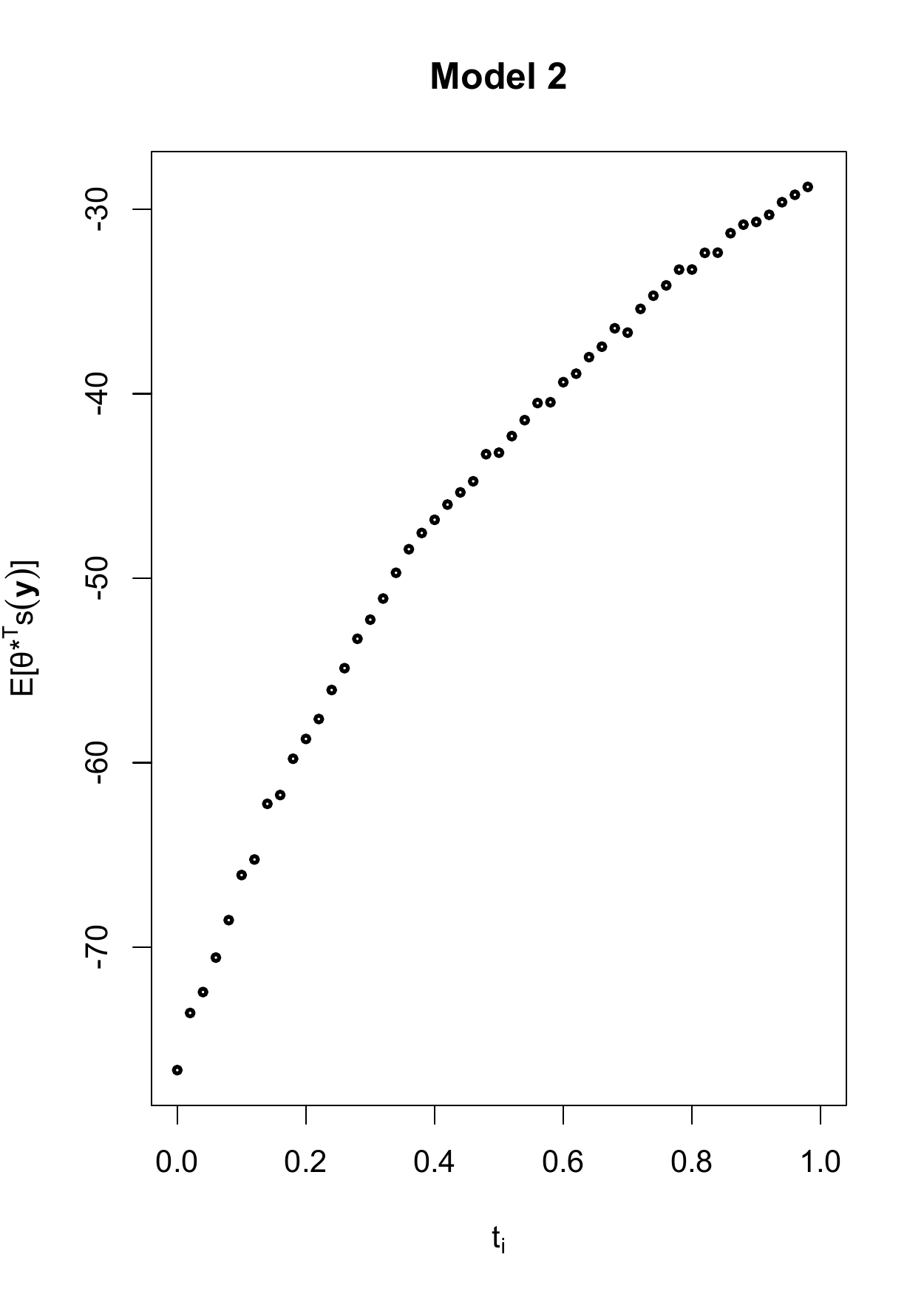}\quad
\includegraphics[scale=0.4]{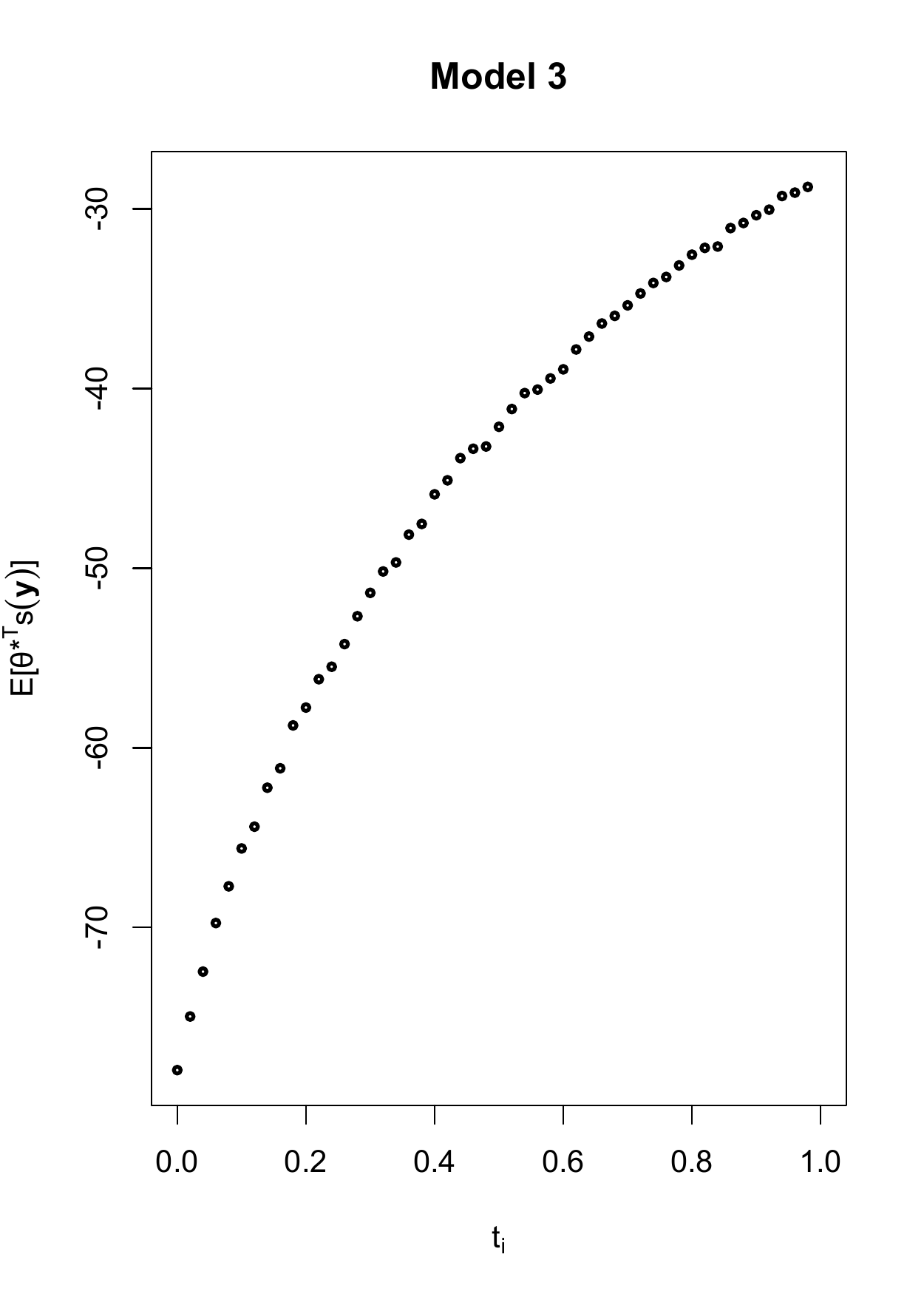}
\caption{$\EE[\bftheta^{\star T} s(\bfy)]$ estimated from a ladder of 50 equally-spaced path points.}
\label{fig:gamaneg_path}
\end{figure}

In terms of calculating the evidence based on path sampling, Figure \ref{fig:gamaneg_path} shows the behaviour of 
$\EE_{\bfy|\bftheta^{\star} t}\left[\bftheta^{\star T} s(\bfy)\right]$ for $50$ equally-spaced path points $t_i$ from 0 to 1. 
The larger the number of temperatures $I$ and the number of simulated networks, the more precise the estimate of the likelihood 
normalizing constant and the greater the computing effort. In this example we estimated (\ref{eq:pathsampling}) using $100$ path 
points and sampling $500$ network statistics for each of them. In this case, this setup has been empirically shown to be 
sufficiently accurate. We set $c$ to be equal to 1 for all the models. However different choices for $c$ do not seem to have a 
big influence on the estimation results if $I$ is large enough. 

A nonparametric density estimation of $p(\bftheta|\bfy)$ for each competing model was implemented using approximate posterior 
samples gathered from the output of the exchange algorithm. Bayes Factor estimates for different sample sizes (which are 
increasing with the number of model dimension) are reported in Table \ref{tab:gamaneg_bf2}. The results are consistent with the 
ones obtained by RJ exchange algorithm displayed in Table \ref{tab:gamaneg_bf}. In particular it is possible to observe that 
as the sample sizes increases the Bayes Factor estimates tend to get closer to the Bayes Factor estimate obtained by the RJ exchange algorithm.
The evidence-based approach took about a few seconds to estimate model evidence for $m_1$ and $m_2$ and about 6 minutes for model $m_3$ using the biggest sample sizes displayed in Table \ref{tab:gamaneg_bf2}.

\begin{table}[htp]
\centering
\begin{tabular}{l|cccc}
\hline\hline
& \multicolumn{4}{c}{Sample sizes}\\
\hline
Model $m_1$ & $100$    & $500$    &  $1,000$     & $5,000$  \\
Model $m_2$ & $150$    & $750$    &  $1,500$     & $7,500$  \\
Model $m_3$ & $200$    & $1,000$  &  $2,000$     & $10,000$ \\
\hline 
$\widehat{BF}_{1,2}$  & $18.83$    & $18.72$    &  $18.84$   & $19.09$   \\
$\widehat{BF}_{1,3}$  & $1029.67$  & $1324.61$  &  $1363.91$   & $1390.08$ \\
\hline\hline
\end{tabular}
\caption{Bayes Factor estimates for increasing values of sample sizes used for the posterior density estimation.}
\label{tab:gamaneg_bf2}
\end{table}

The estimates of the Bayes Factors can be interpreted using the guidelines of \cite{kas:raf95}, Table~\ref{tab:kas_raf}, leading
to the conclusion that the Bayes Factor estimates obtained suggest 
that there is positive/strong evidence in favour of model $m_1$ which is the one including the number of edges against the 
other two competing models. Thus in this case the only strong effect of the antagonistic structure of the Gahuku-Gama 
tribes is represented by the low edge density.

\subsubsection{Gamapos}

\begin{figure}[htp]
\centering
\includegraphics[scale=0.5]{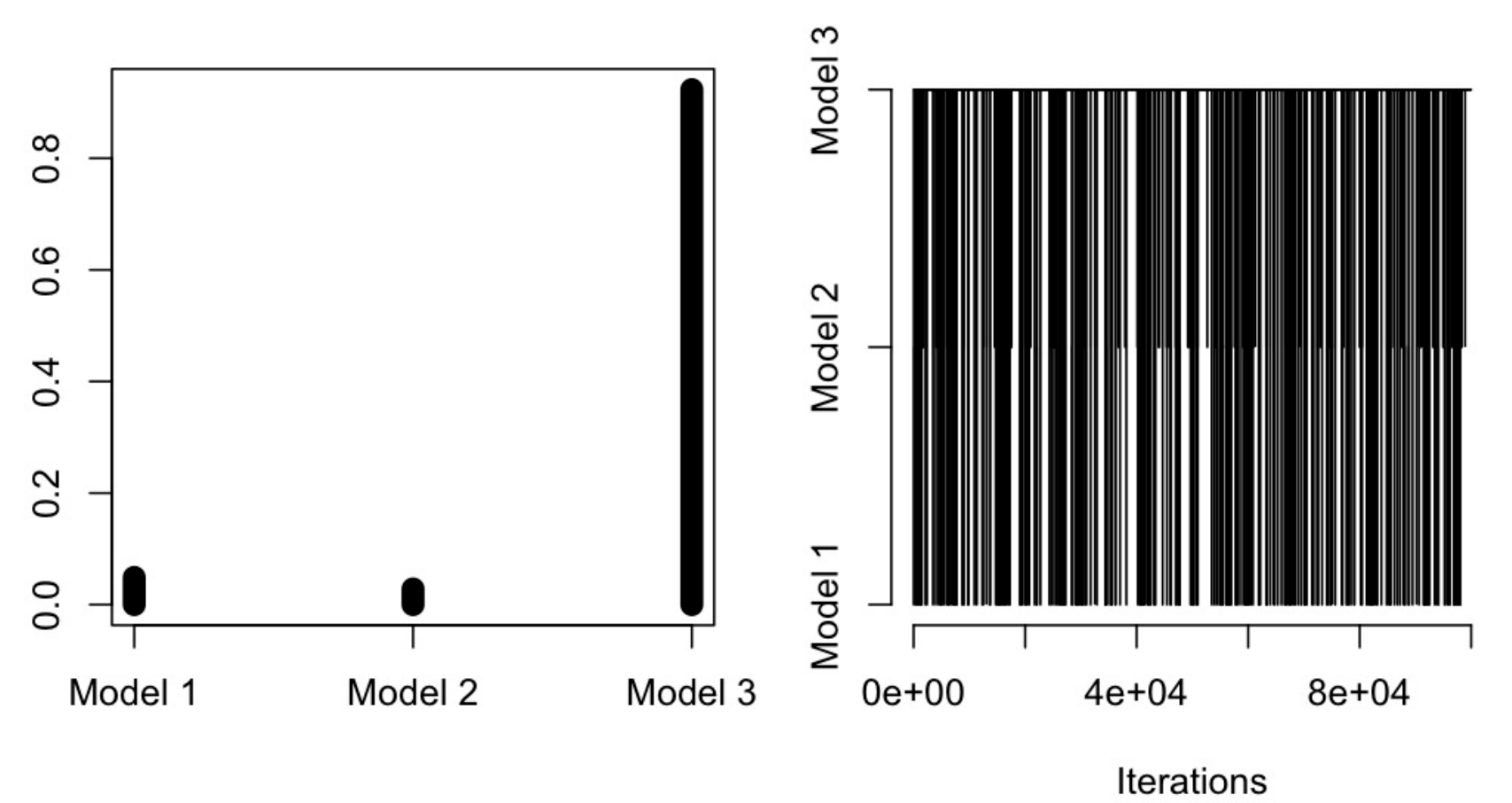}\\
\vspace{1cm}
\includegraphics[scale=0.9]{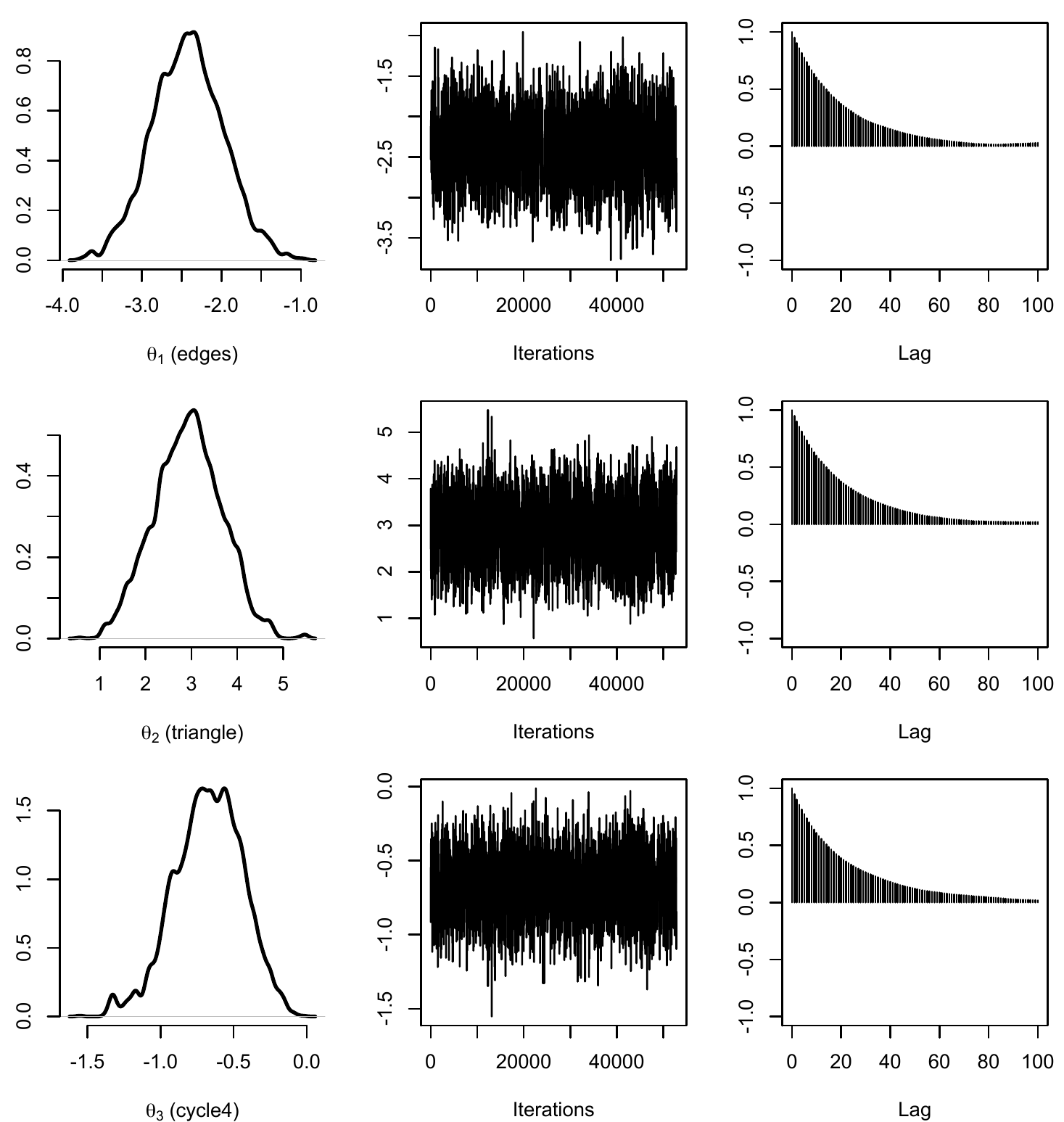}
\caption{Auto-RJ exchange algorithm output: posterior model probabilities (top) and posterior parameter probabilities for model $m_3$ (bottom).}
\label{fig:gamapos_post}
\end{figure}

In this second example, we considered the same competing models of Table \ref{tab:gamaneg_modprop}. In this case it turned out that the pilot-tuned RJ exchange algorithm was very difficult to tune, being very sensitive to the choice of the parameters of jump proposal. 
We used the auto-RJ exchange algorithm with the same set-up of the previous example. The output from auto-RJ exchange algorithm is displayed in Figure \ref{fig:gamapos_post} and the parameter posterior estimates in Table \ref{tab:gamapos_parpost}.

\begin{table}[htp]
\centering
\begin{tabular}{l|cc}
\hline\hline
Parameter & Post. Mean & Post. Sd.\\
\hline
\multicolumn{3}{c}{Model $m_3$ (within-model acc. rate: $0.3$)}\\
\hline
$\theta_1$ (edge)      & -2.41 & 0.45\\
$\theta_2$ (triangle)  &  2.91 & 0.71\\
$\theta_3$ (4-cycle)   & -0.66 & 0.22\\
\hline
\multicolumn{3}{c}{Model $m_1$ (within-model acc. rate: $0.64$)}\\
\hline
$\theta_1$ (edge)      & -1.15 & 0.20\\
\hline
\multicolumn{3}{c}{Model $m_2$ (within-model acc. rate: $0.3$)}\\
\hline
$\theta_1$ (edge)      & -1.69 & 0.35\\
$\theta_2$ (triangle)  &  0.48 & 0.20\\
\hline
\multicolumn{3}{c}{Between-model acc. rate: $0.03$}\\
\hline\hline
\end{tabular}
\caption{Summary of posterior parameter estimates and acceptance rates.}
\label{tab:gamapos_parpost}
\end{table}

\begin{figure}[htp]
\centering
\includegraphics[scale=0.36]{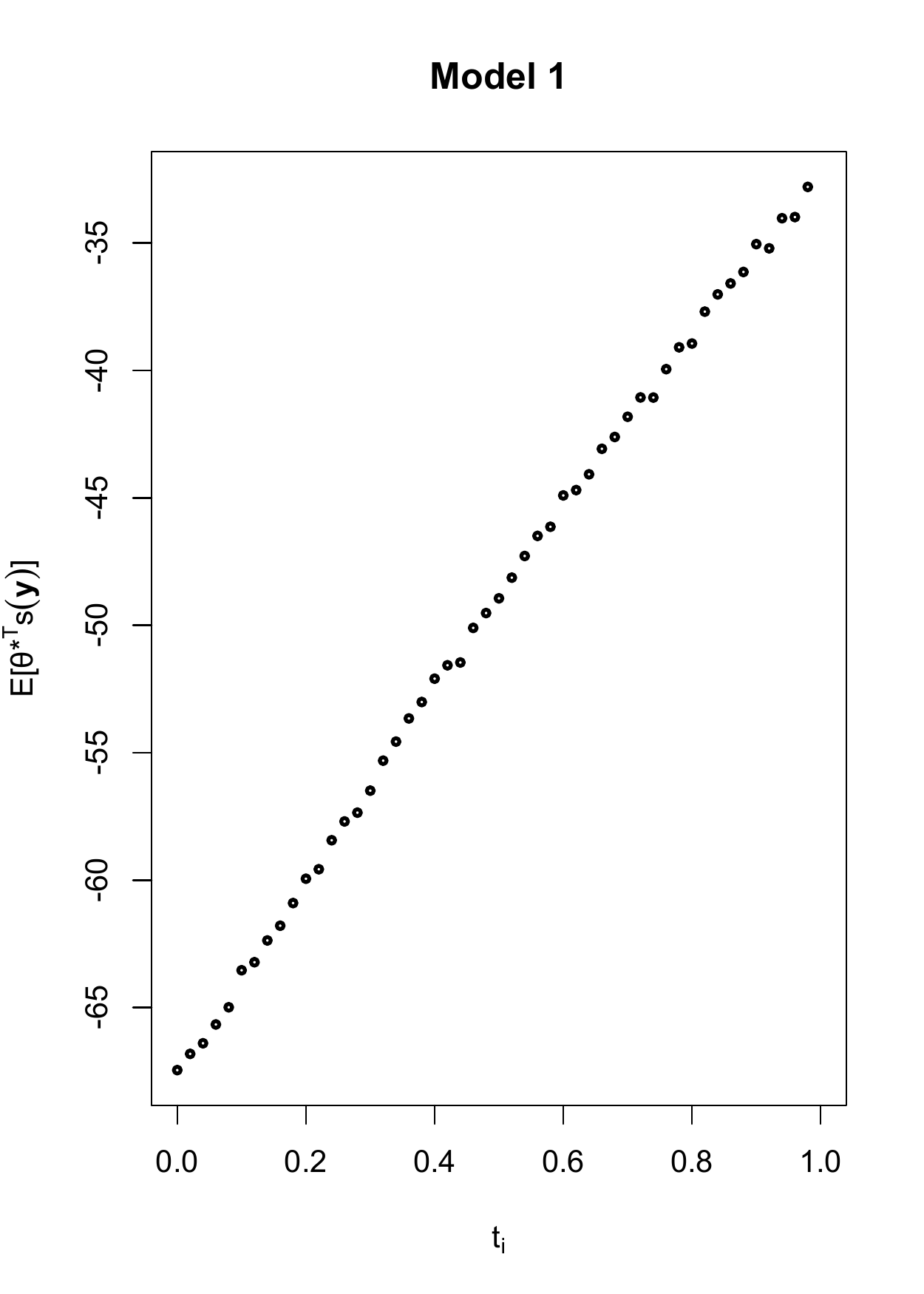}
\includegraphics[scale=0.36]{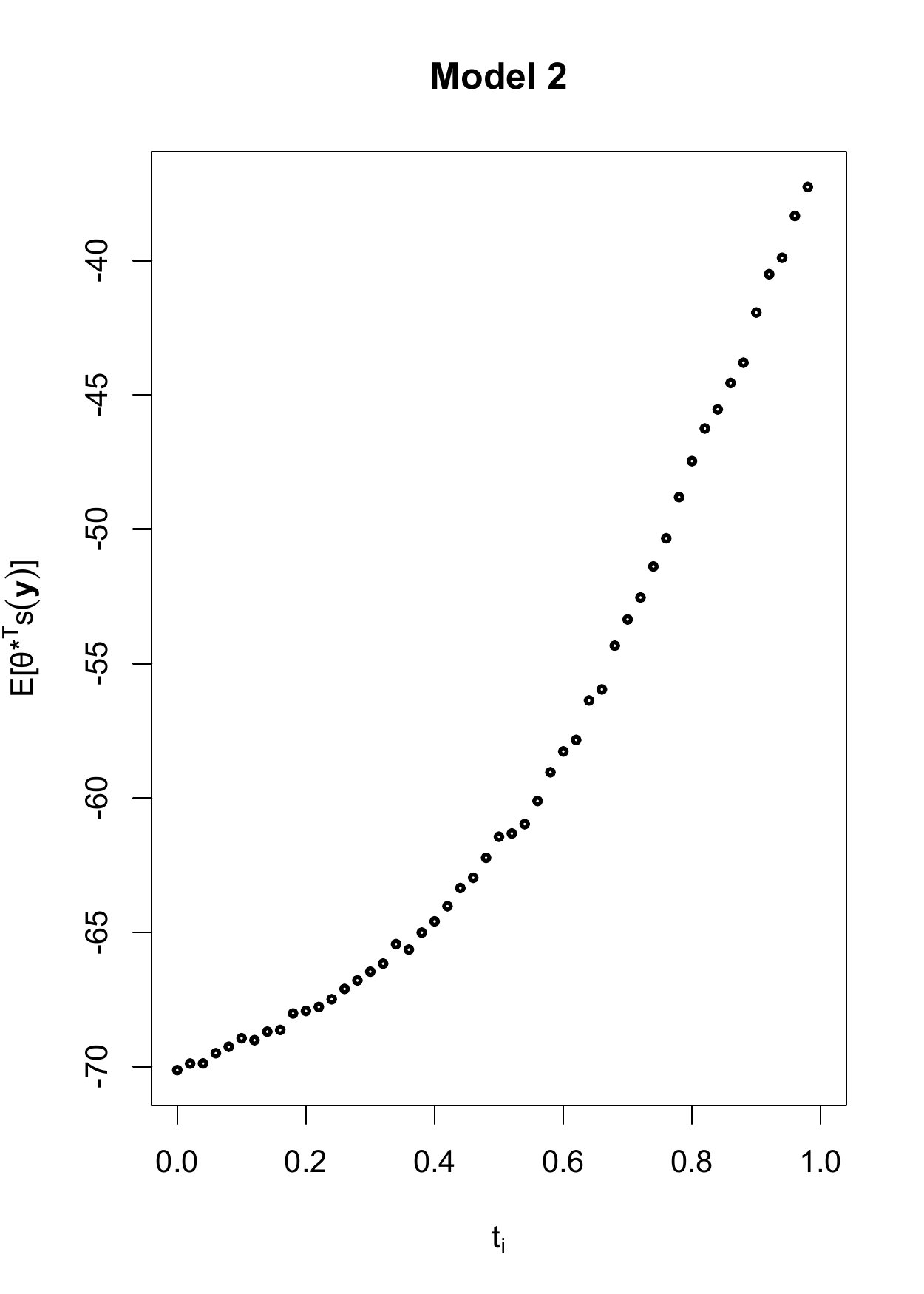}
\includegraphics[scale=0.36]{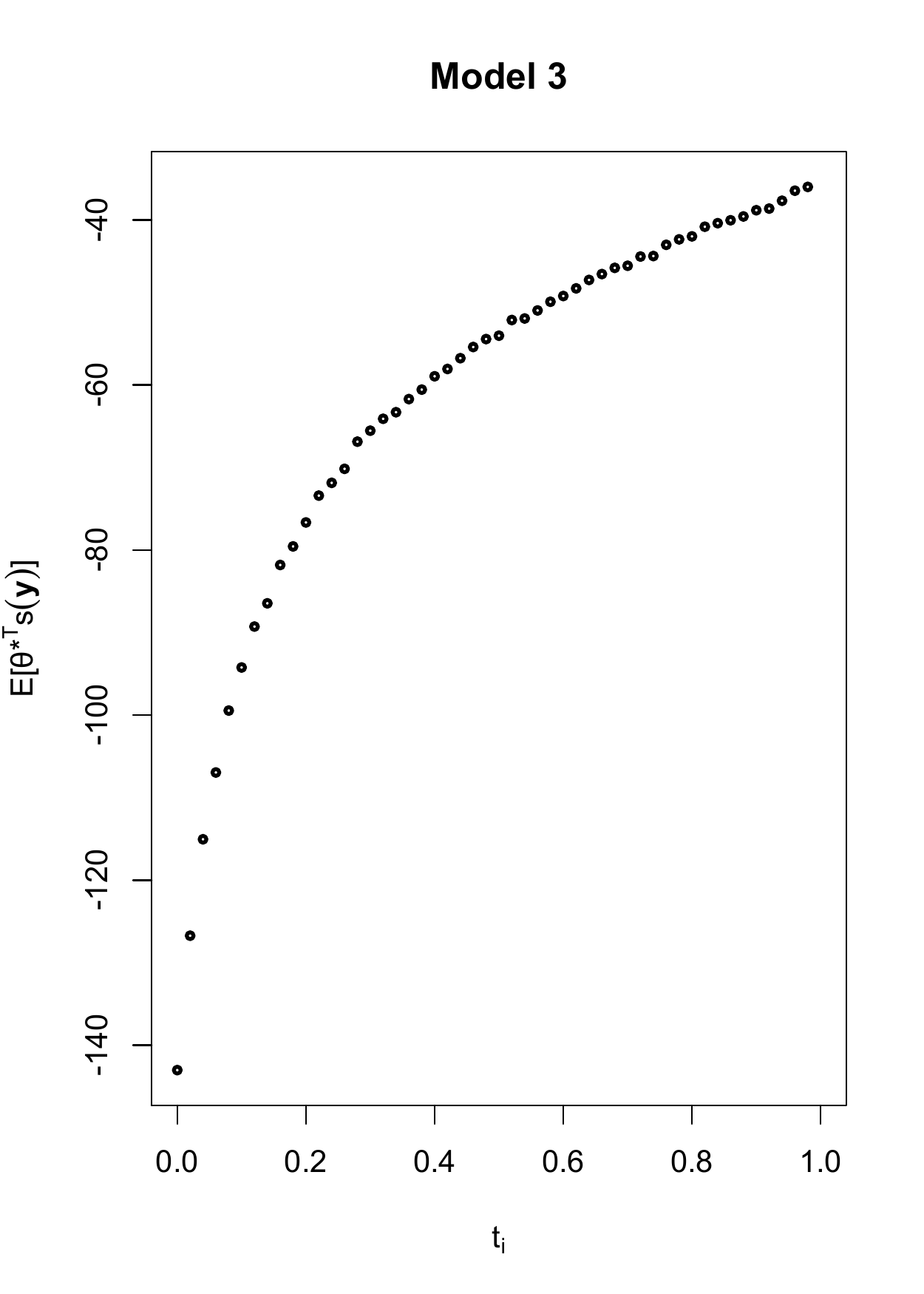}
\caption{$\EE[\bftheta^{\star T} s(\bfy)]$ estimated from a ladder of 50 equally-spaced path points.}
\label{fig:gamapos_path}
\end{figure}

We also calculated the evidence for each models following the same setup of the Gamaneg example. Figure~\ref{fig:gamapos_path} shows the behaviour of 
$\EE_{\bfy|\bftheta^{\star} t}\left[\bftheta^{\star T} s(\bfy)\right]$ for $50$ equally-spaced path points $t_i$ from 0 to 1.  
Table~\ref{tab:gamapos_bf} reports the Bayes Factor estimates of the auto-RJ exchange algorithm and evidence-based method using 
the biggest sample sizes used for the posterior density estimation of the previous example. From this one can conclude that there
is positive/strong support for model $m_3$.

\begin{table}[htp]
\centering
\begin{tabular}{l|c|c}
\hline\hline
& Auto-RJ algorithm & Evidence-based method\\
\hline
$BF_{3,1}$ & $17.83$   & $19.31$\\
$BF_{3,2}$ & $34.81$   & $32.82$\\
\hline\hline
\end{tabular}
\caption{Bayes factors estimates.}
\label{tab:gamapos_bf}
\end{table}

In the Gamapos network the transitivity and the 4-cycle structure are important features of the network. The tendency to a low 
density of edges and 4-cycles expressed by the negative posterior mean of the first and third parameters is balanced by a 
propensity for local triangles which gives rise to the formation of small well-defined alliances.

We remark that both examples should be considered from a pedagogical viewpoint, and not from a solely applied perspective. However 
it is interesting that although both networks are defined on the same node set, the model selection procedures for each example lead 
to different models having highest probability, a posteriori. 
% DEGENERACY ISSUE OF MODEL 2 %
It is also important to note that model $m_2$ is known to be a degenerate model (see \cite{jon99}, \cite{but11}, and \cite{sha:rin11}) 
as it tends to place almost all probability mass on extreme graphs under almost all values of the parameters. For this reason 
model $m_2$ is unrealistic for real-world networks. Indeed, it may be suspected that model $m_3$ is potentially problematic, however
the asymptotic properties of this model has not yet been studied. Our Bayesian model choice procedures agree with the previous
knowledge of $m_2$, as outlined above, in the sense 
that very little posterior probability is assigned to model $m_2$ in both the examples above. One may view this as a useful check of 
the reliability of the algorithm.

\subsection{Collaboration between Lazega's lawyers}

\begin{figure}[htp]
\centering
\includegraphics[scale=.5]{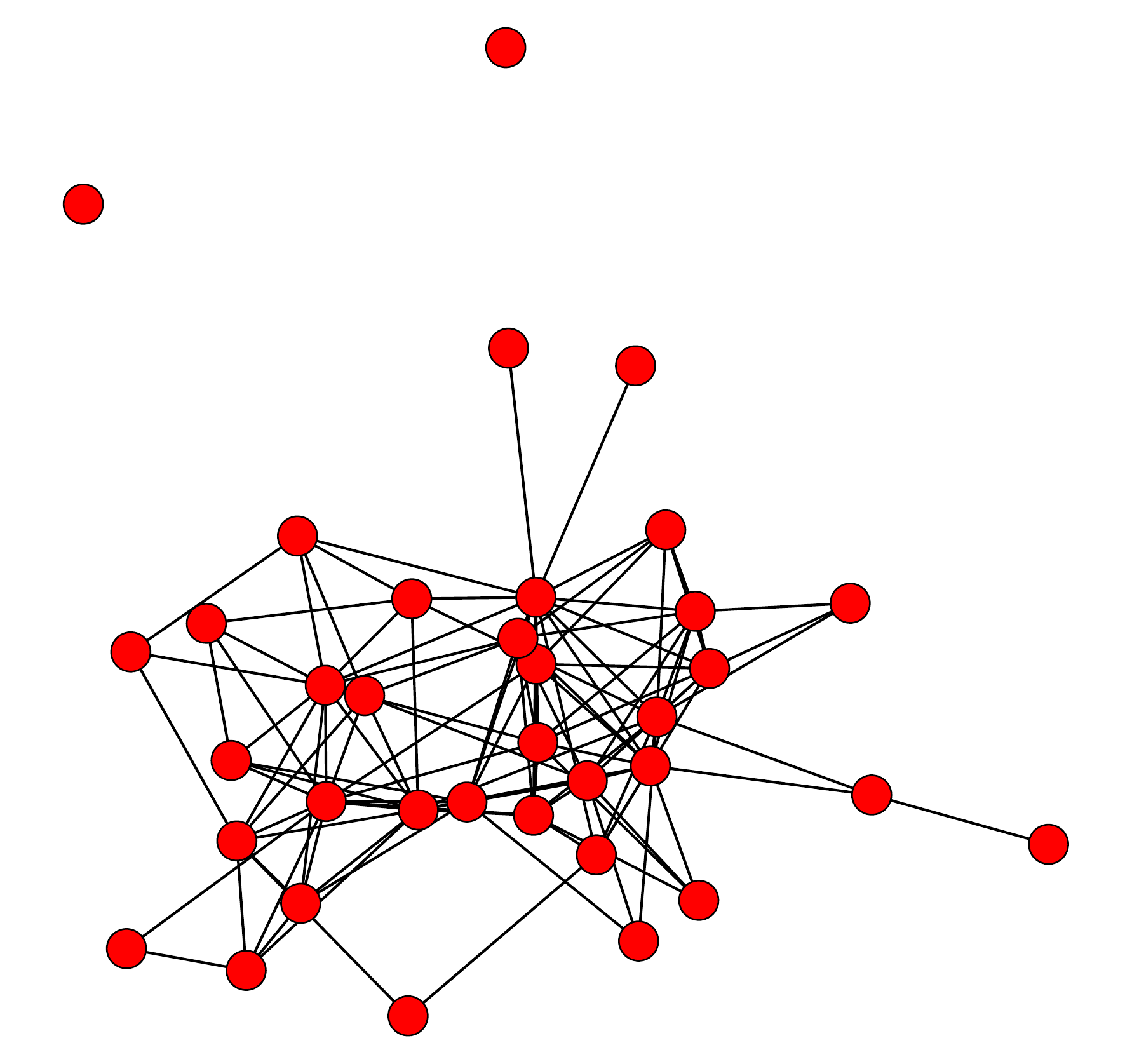}
\caption{Lazega's lawyers cowork graph.}
\label{fig:lazega_graph}
\end{figure}

The Lazega network data collected by \cite{laz01} and displayed in Figure \ref{fig:lazega_graph} represents the symmetrized collaboration relations between the $36$ partners in a New England law firm, where the presence of an edge between two nodes indicates that both partners collaborate with the other.

\subsubsection{Example 1}

In this example we want to compare 4 models (Table \ref{tab:lazega_modprop}) using the edges, geometrically weighted degrees and geometrically weighted edgewise shared partners \citep{sni:pat:rob:han06}:
\begin{center}
\begin{tabular}{ll}
edges & $\sum_{i<j}y_{ij}$\\
geometrically weighted degree (gwd) & $e^{\phi_u} \sum_{k=1}^{n-1} 
\left\{ 1- \left( 1 - e^{-\phi_u} \right )^{k} \right \} D_k(\bfy)$ \\
geometrically weighted edgewise & $
e^{\phi_v} \sum_{k=1}^{n-2}
\left \{ 1-\left( 1 - e^{-\phi_v} \right)^{k} \right \} EP_k(\bfy)$ 
\\
shared partner (gwesp) &
\end{tabular}
\end{center}
where $\phi_u=\log(2)$, $\phi_v=\log(2)$, $D_k(\bfy)$ is the number of pairs that have exactly $k$ common neighbours and $EP_k(\bfy)$ is the number of connected pairs with exactly $k$ common neighbours. 

\begin{table}[htp]
\centering
\begin{tabular}{ll}
\hline\hline
Model $m_1$ & $\bfy \sim$ edges\\
Model $m_2$ & $\bfy \sim$ edges $+$ gwesp($\log(2)$)\\
Model $m_3$ & $\bfy \sim$ edges $+$ gwesp($\log(2)$) $+$ gwd($\log(2)$)\\
Model $m_4$ & $\bfy \sim$ edges $+$ gwd($\log(2)$)\\
\hline\hline
\end{tabular}
\caption{Competing models.}
\label{tab:lazega_modprop}
\end{table}

\begin{figure}[htp]
\centering
\includegraphics[scale=0.5]{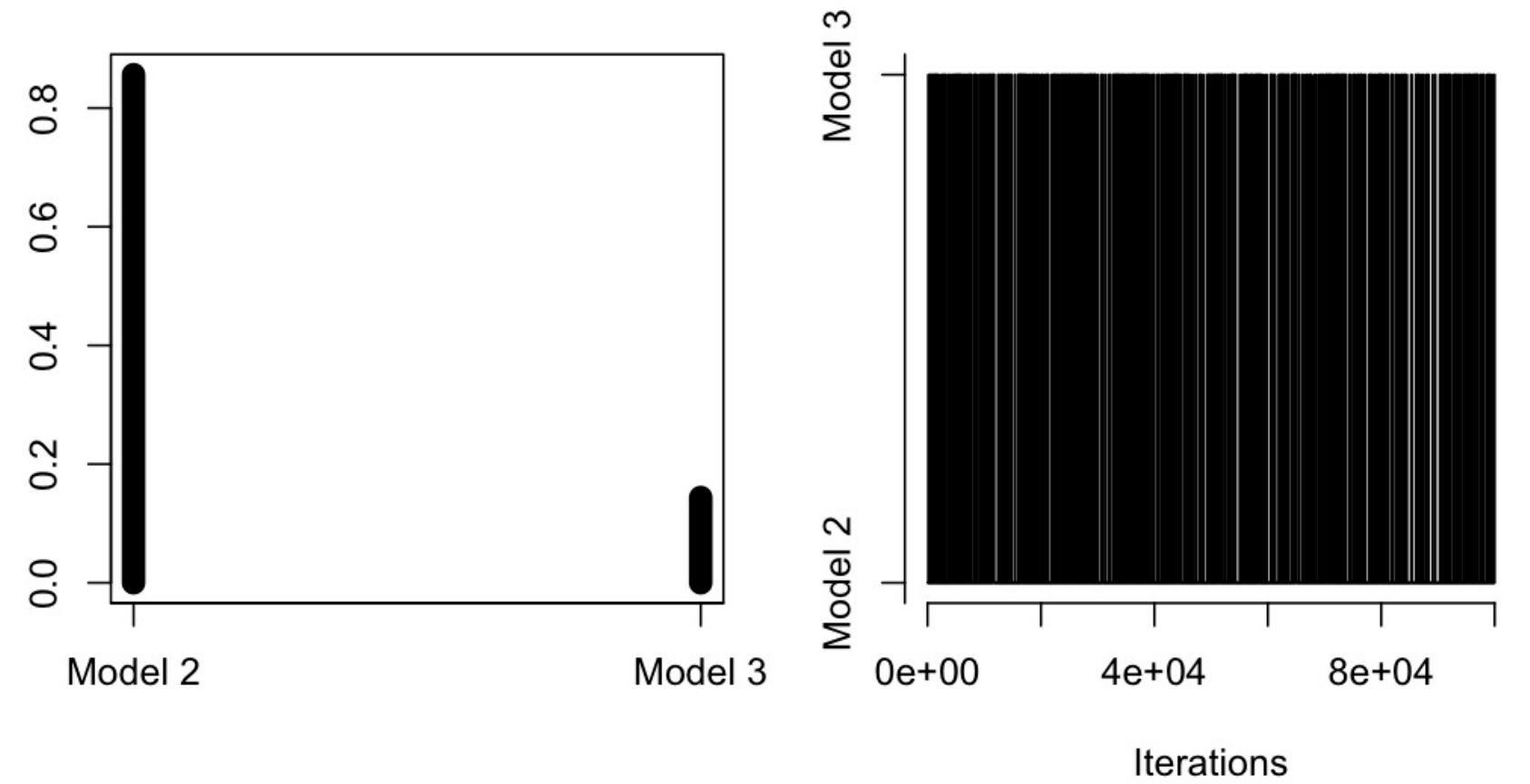}\\
\vspace{1cm}
\includegraphics[scale=0.9]{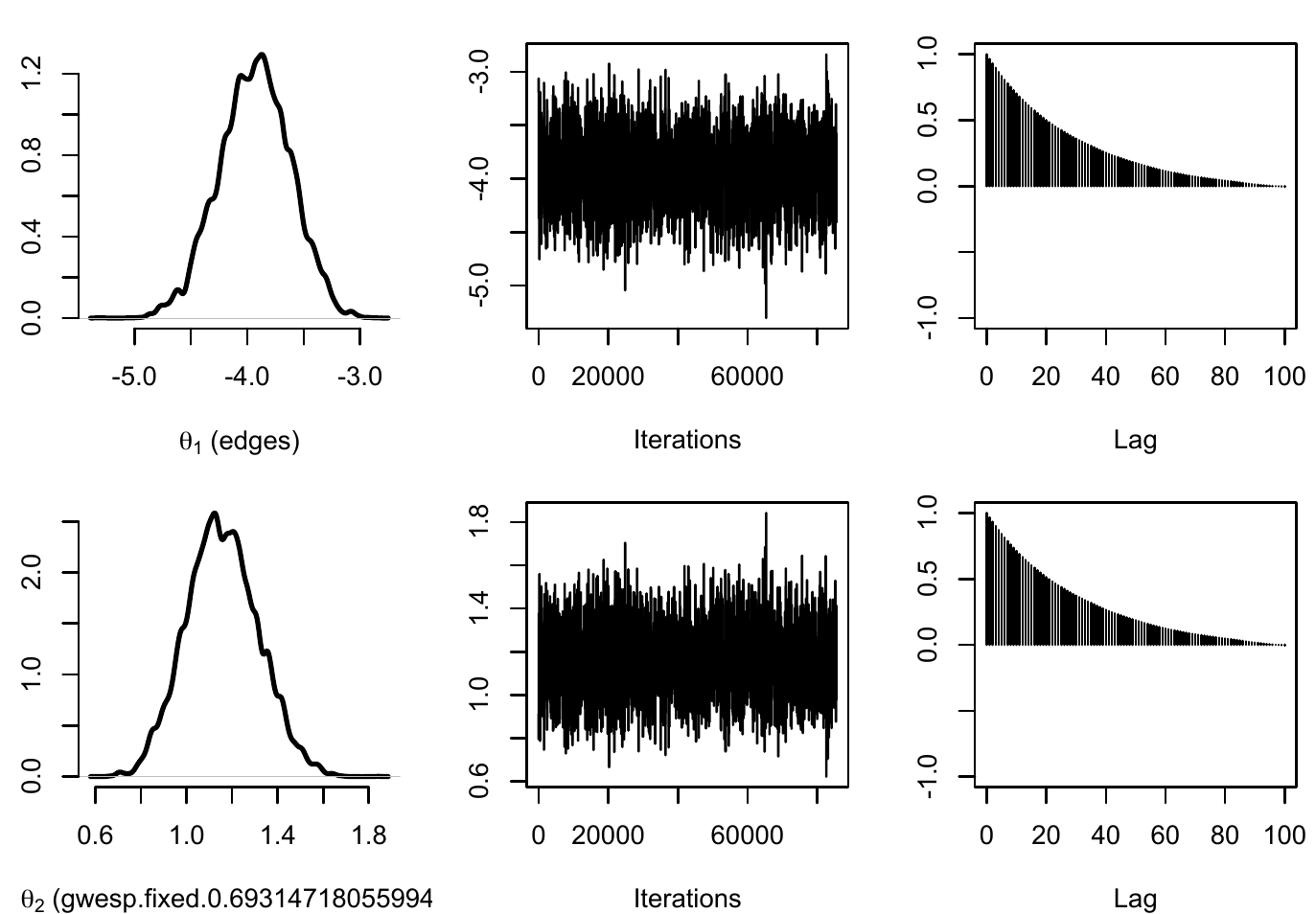}
\caption{Auto-RJ exchange algorithm output: posterior model probabilities (top) and posterior parameter probabilities for model $m_2$ (bottom).}
\label{fig:lazega_post}
\end{figure}

As happened in the previous example, the pilot-tuned RJ exchange algorithm proved to be ineffective due to the difficulty of the 
tuning problem. The auto-RJ exchange algorithm was run for $100,000$ iterations using the same flat normal priors of the previous 
examples and $25,000$ auxiliary iterations for network simulation.
The offline run consisted of estimating $\hat{\bfmu}_{l}$ and $\hat{\bfSigma}_{l}$ for each of the 4 models by using 
$6,000 \times D_{l}$ main iterations (discarding the first $1,000 \times D_{l}$ iterations as burnin). The algorithm took about 
1 hour and 50 minutes to complete the estimation, the results of which are displayed in Figure \ref{fig:lazega_post} and 
Table \ref{tab:lazega_parpost}.

\begin{table}[htp]
\centering
\begin{tabular}{l|cc}
\hline\hline
Parameter & Post. Mean & Post. Sd.\\
\hline
\multicolumn{3}{c}{Model $m_2$ (within-model acc. rate: $0.24$)}\\
\hline
$\theta_1$ (edge)          & -3.93 & 0.33\\
$\theta_2$ (gwesp($\log(2)$))  &  1.15 & 0.16\\
\hline
\multicolumn{3}{c}{Model $m_3$ (within-model acc. rate: $0.26$)}\\
\hline
$\theta_1$ (edge)             & -4.54 & 0.56\\
$\theta_2$ (gwesp($\log(2)$))  & -1.39 & 0.23\\
$\theta_3$ (gwd($\log(2)$))    &  0.79 & 0.62\\
\hline
%\multicolumn{3}{c}{Model $m_4$ (within-model acc. rate: $0.38$)}\\
%\hline
%$\theta_1$ (edge)          & -1.21 & 0.12\\
%$\theta_2$ (gwd($0.8$))    & -2.23 & 0.25\\
%\hline
%\multicolumn{3}{c}{Model $m_1$ (within-model acc. rate: $0$)}\\
%\hline
%$\theta_1$ (edge)          & $NA$ & $NA$\\
%\hline
\multicolumn{3}{c}{Between-model acc. rate: $0.03$}\\
\hline\hline
\end{tabular}
\caption{Summary of posterior parameter estimates and acceptance rates.}
\label{tab:lazega_parpost}
\end{table}

The evidence-based algorithm was carried out using $200$ path points from each of which we sampled $500$ networks. The results are reported in Table \ref{tab:lazega_bf}. The algorithm took 25 seconds to estimate the evidence for model $m_1$, 8 minutes for model $m_2$, 9 minutes for model $m_3$, 1 minute for model $m_4$.

\begin{figure}[htp]
\centering
\includegraphics[scale=0.4]{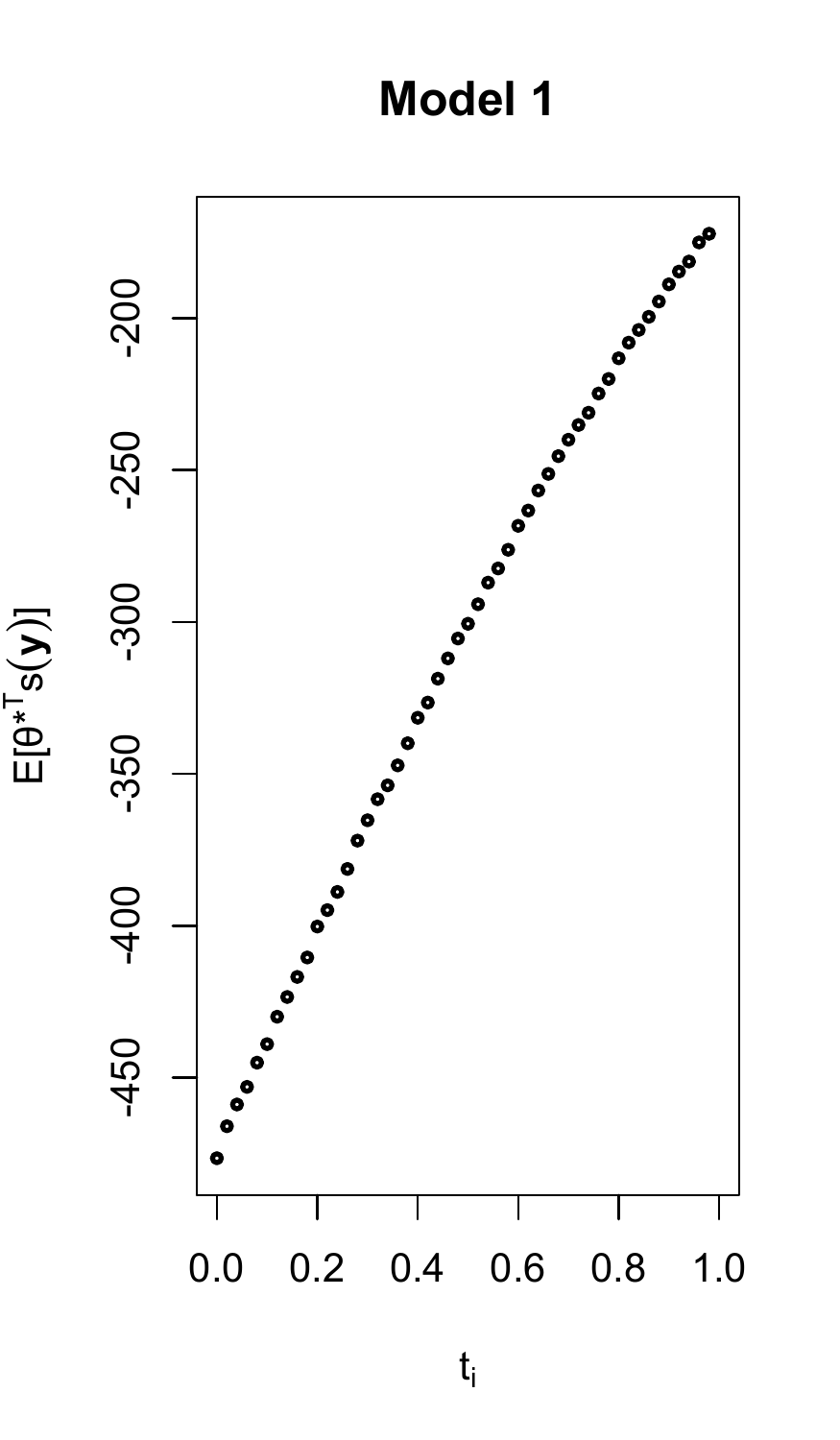}
\includegraphics[scale=0.4]{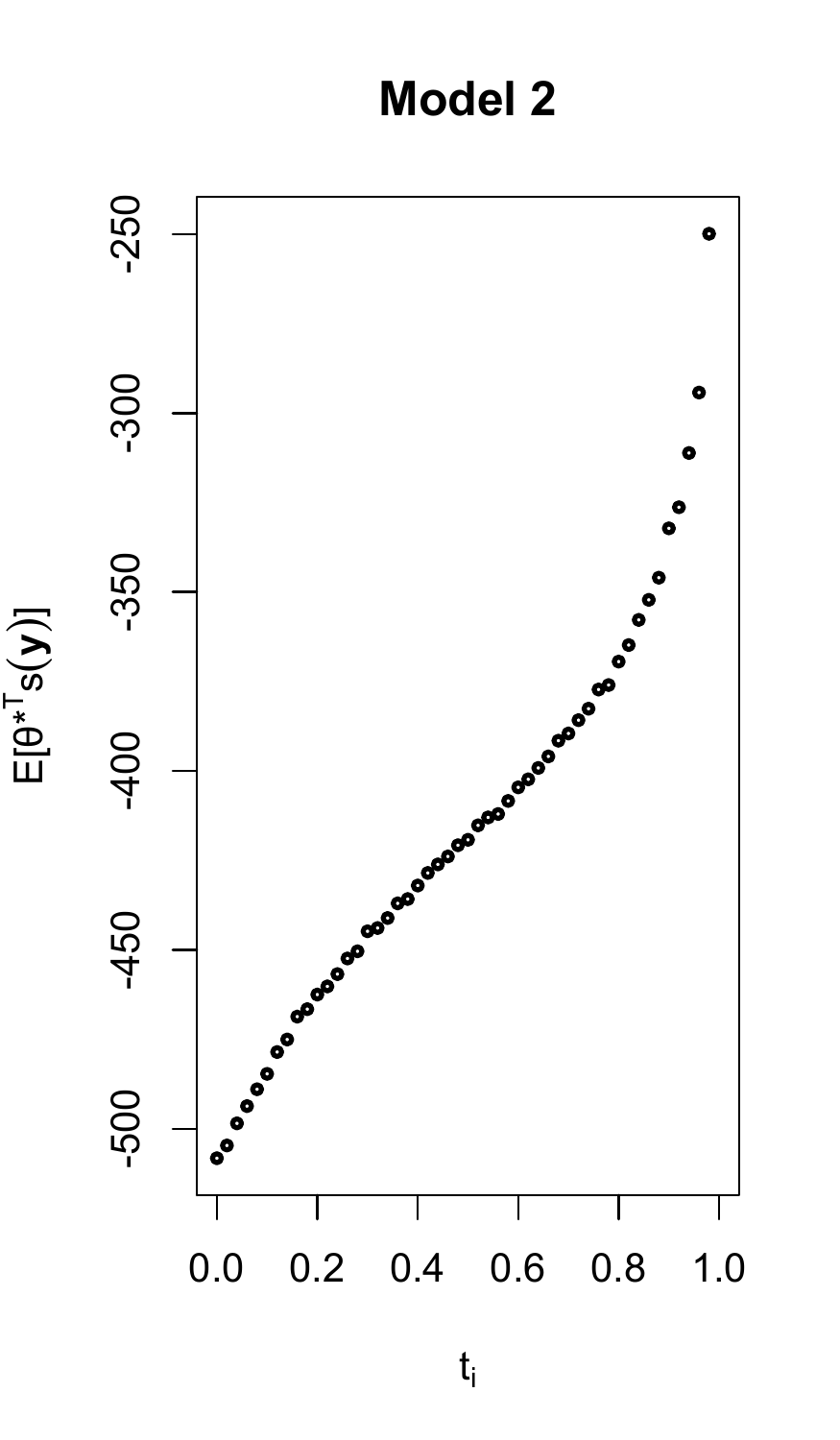}
\includegraphics[scale=0.4]{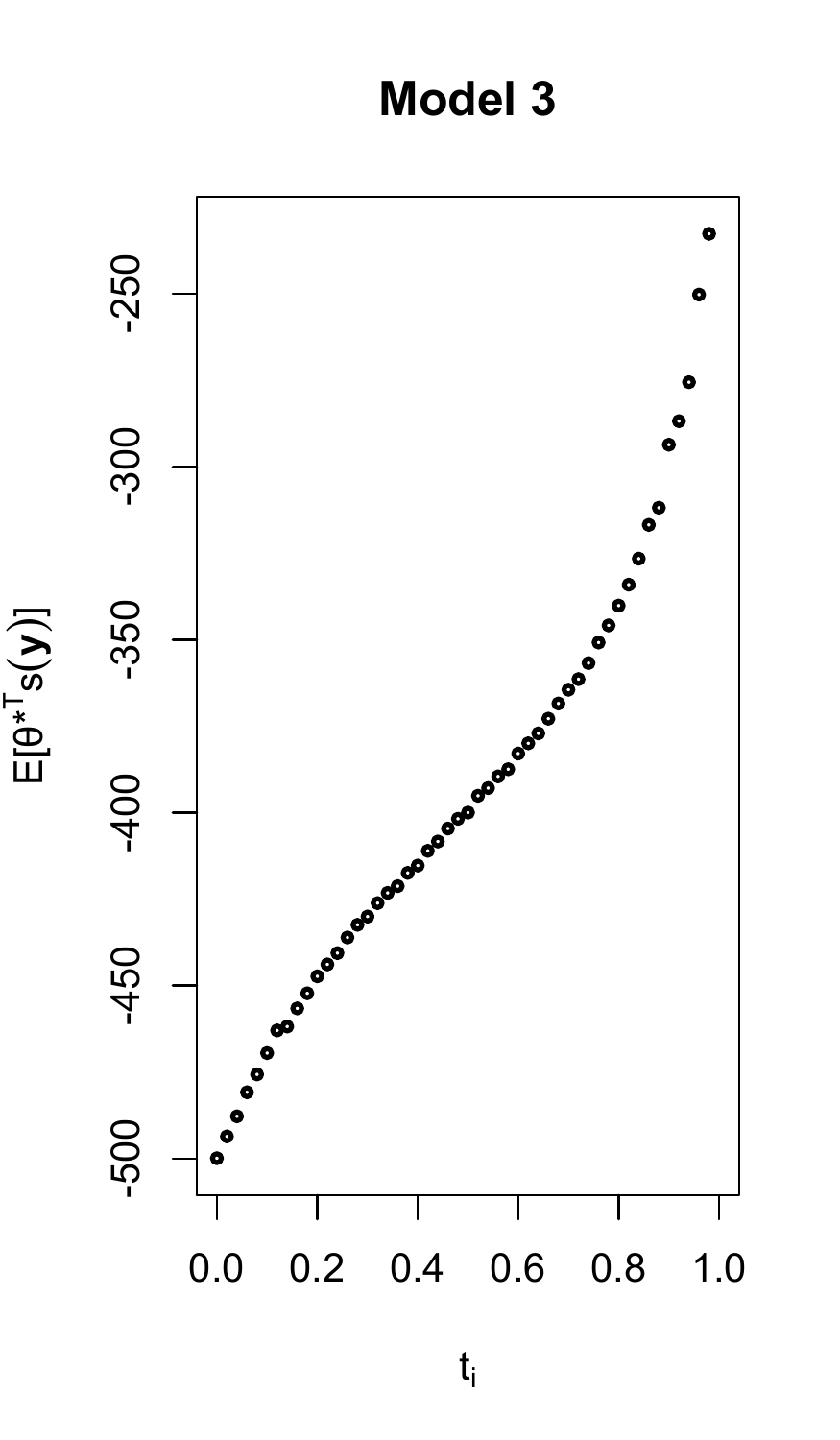}
\includegraphics[scale=0.4]{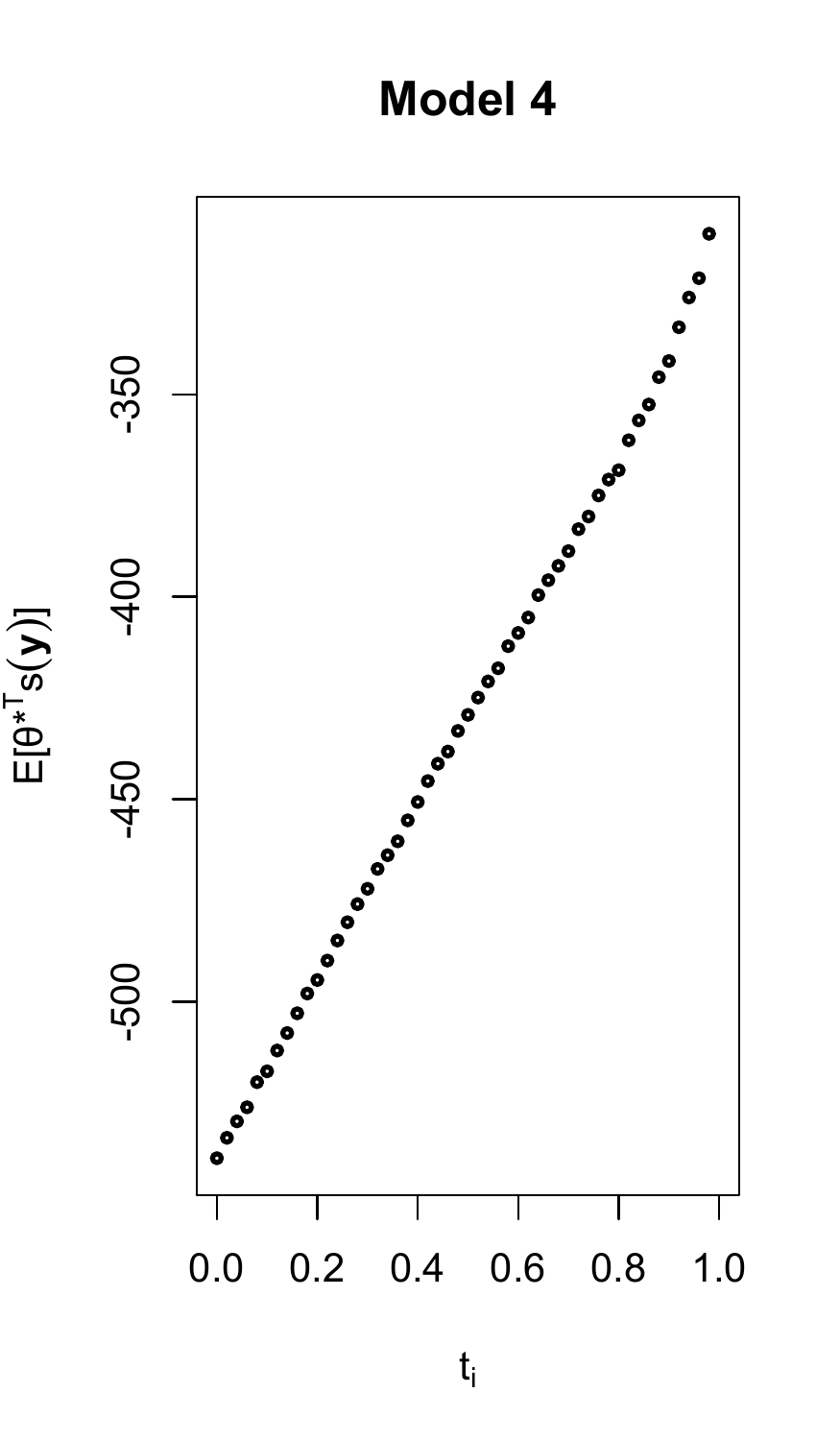}
\caption{$\EE[\bftheta^{\star T} s(\bfy)]$ estimated from a ladder of 50 equally-spaced path points.}
\label{fig:lazega_path}
\end{figure}

\begin{table}[htp]
\centering
\begin{tabular}{l|c|c}
\hline\hline
& Auto-RJ algorithm & Evidence-based method\\
\hline
$BF_{2,1}$ & $>10^{6}$ & $>10^6$\\
$BF_{2,3}$ & $5.72$    & $4.65$\\
$BF_{2,4}$ & $>10^{6}$ & $>10^6$\\
\hline\hline
\end{tabular}
\caption{Bayes Factor estimates.}
\label{tab:lazega_bf}
\end{table}

Table \ref{tab:lazega_bf} displays the Bayes Factor for the comparison between model $m_2$ (best model) against the others. There is positive evidence to reject model $m_3$ and very strong evidence to models $m_1$ and $m_4$.

We can therefore conclude that the low density effect expressed by the negative edge parameter combined with the positive transitivity effect expressed by the geometrically weighted edgewise partners parameter are strong structural features not depending on popularity effect expressed by the weighted degrees. These results are in agreement with the findings reported in the literature (see \cite{sni:pat:rob:han06} and \cite{hun:han06}).
However, the advantage of the Bayesian approach used in this paper is that the comparison between competing models is carried out within a fully probabilistic framework while classical approaches test the significativity of each parameter estimate using t-ratios defined as parameter estimate divided by standard error, and referring these to an approximating standard normal distribution as the null distribution.

\subsubsection{Example 2}

In this example we want to compare the two models specified in Table~\ref{tab:lazega2_modprop} using the edges, geometrically 
weighted edgewise shared partners (with $\phi_v=\log(2)$) and a set of statistics involving exogenous data based on some nodal 
attributes available in the Lazega dataset. In particular we consider the following nodal covariates: gender and practice 
(2 possible values, litigation$=0$ and corporate law$=1$). The covariate statistics are of the form:
\begin{equation*}
s(\bfy,\bfx) = \sum_{i\neq j} y_{ij}f(x_i,x_j)
\end{equation*}
where $f(x_i,x_j)$ can either define a ``main effect'' of a numeric covariate:
\begin{equation*}
f(x_i,x_j) = x_{i} + x_{j}
\end{equation*}
or a ``similarity effect'' (or ``homophily effect''):
\begin{equation*}
f(x_i,x_j) = \bfI_{\{x_{i} = x_{j}\}}
\end{equation*}
where $\bfI$ is the indicator function.
\begin{table}[htp]
\centering
\begin{tabular}{l|l}
\hline\hline
Model $m_1$ & Model $m_2$\\
\hline
edges & edges\\
gwesp($\log(2)$) & gwesp($\log(2)$)\\
practice - homophily & gender - homophily\\
law-school - homophily & practice - homophily\\
practice - main effect & \\
\hline\hline
\end{tabular}
\caption{Competing models.}
\label{tab:lazega2_modprop}
\end{table}

\begin{figure}[htp]
\centering
\includegraphics[scale=0.5]{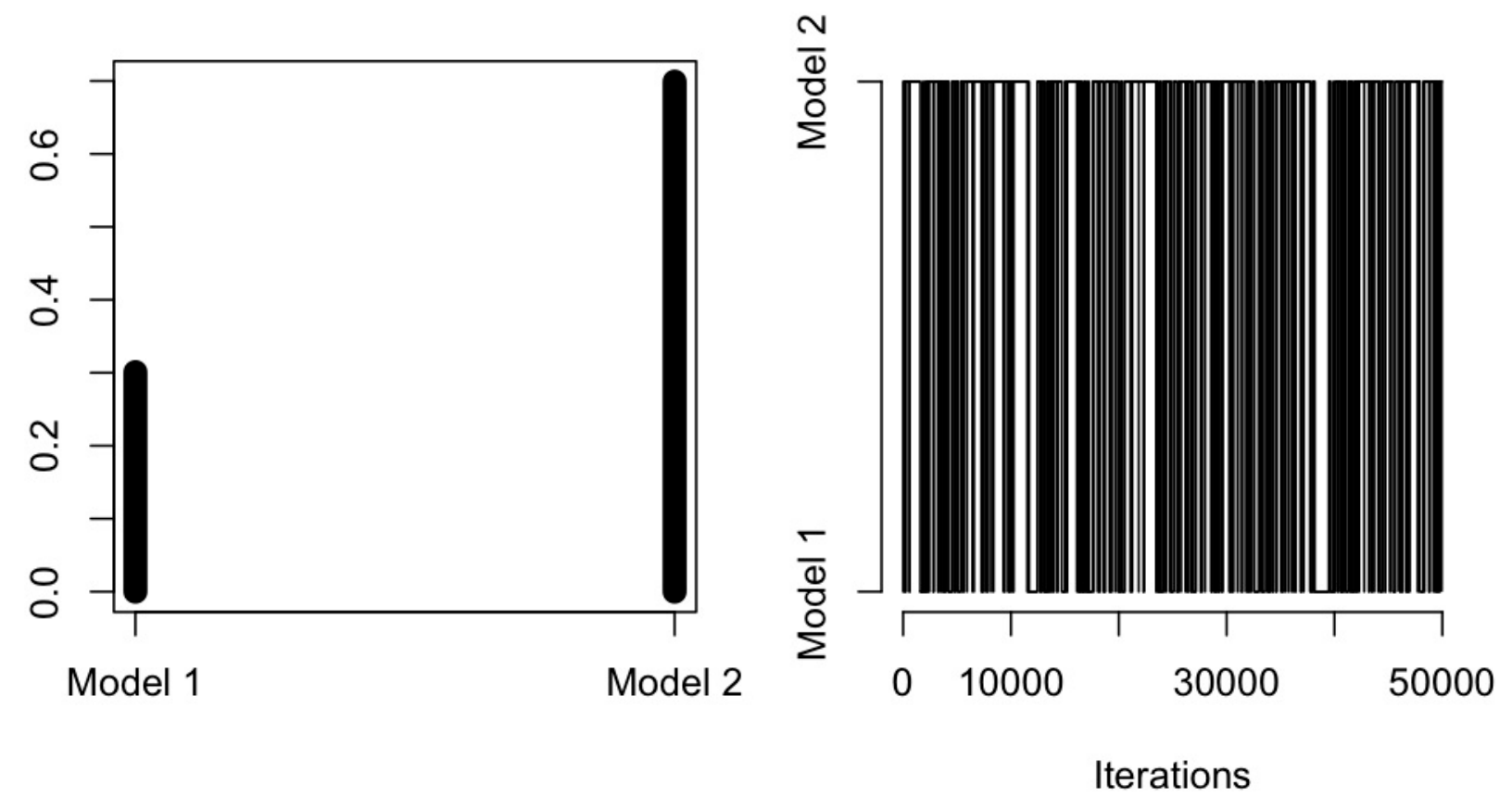}\\
\vspace{1cm}
\includegraphics[scale=1]{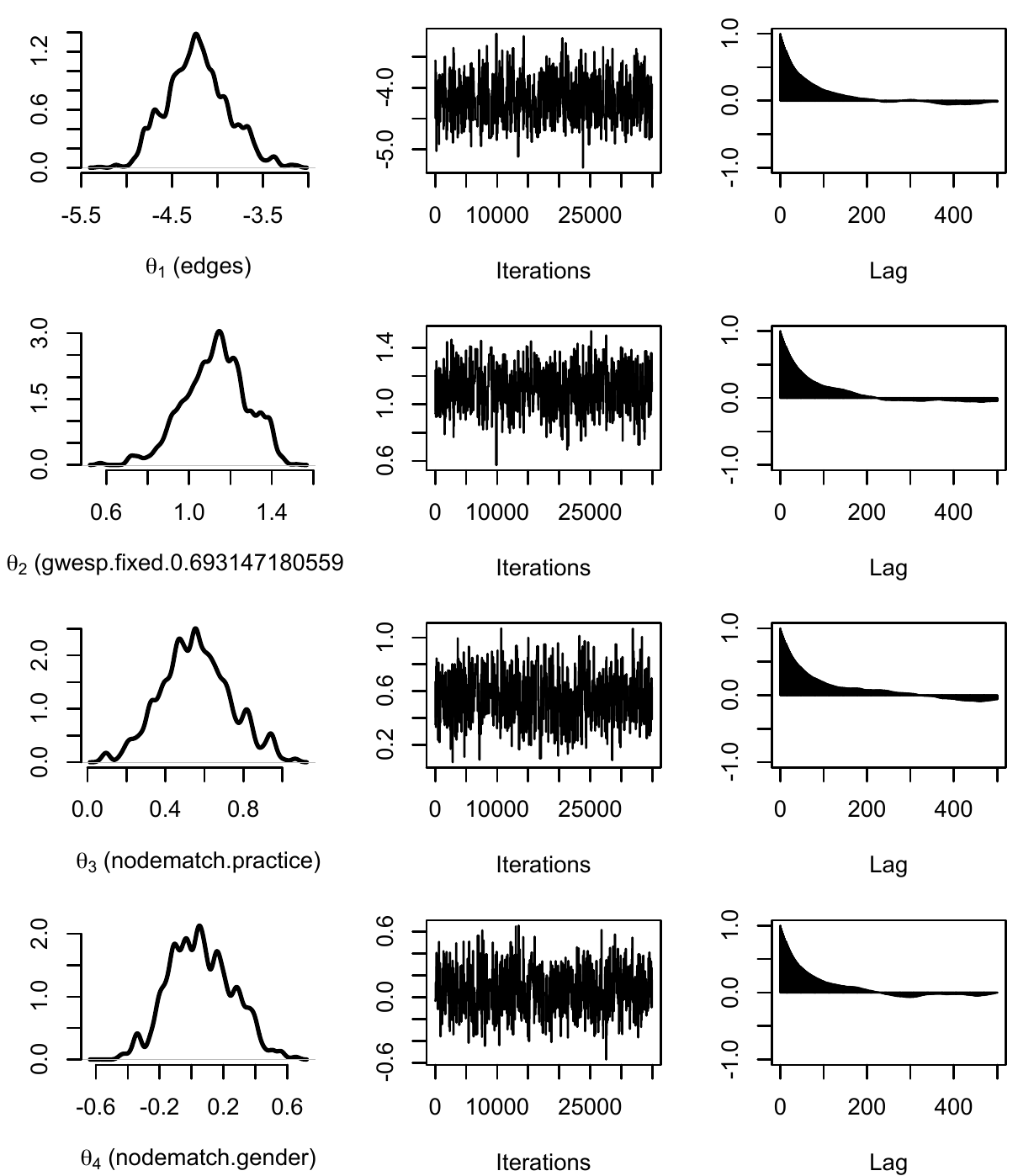}
\caption{Auto-RJ exchange algorithm output: posterior model probabilities (top) and posterior parameter probabilities for model $m_2$ (bottom).}
\label{fig:lazega2_post}
\end{figure}

In this case, due to the high-dimensionality of both the competing models, only the auto-RJ exchange approach was used. The algorithm was run for $50,000$ iterations using the same flat normal priors of the previous examples and $25,000$ auxiliary iterations for network simulation.
The offline run consisted of estimating $\hat{\bfmu}_{l}$ and $\hat{\bfSigma}_{l}$ for each of the 2 models by using 
$5,000 \times D_{l}$ main iterations (discarding the first $1,000 \times D_{l}$ iterations as burnin). The algorithm took about 2 hours to complete the estimation, the results of which are displayed in Figure \ref{fig:lazega2_post} and 
Table \ref{tab:lazega2_parpost}.

\begin{table}[htp]
\centering
\begin{tabular}{l|cc}
\hline\hline
Parameter & Post. Mean & Post. Sd.\\
\hline
\multicolumn{3}{c}{Model $m_2$ (within-model acc. rate: $0.11$)}\\
\hline
$\theta_1$ (edge)                  & $-4.22$ & $0.34$\\
$\theta_2$ (gwesp(log(2)))         & $1.12$ & $0.15$\\
$\theta_3$ (practice - homophily)  & $0.55$ & $0.17$\\
$\theta_4$ (gender - homophily)    & $0.05$ & $0.19$\\
\hline
\multicolumn{3}{c}{Model $m_1$ (within-model acc. rate: $0.13$)}\\
\hline
$\theta_1$ (edge)                    & $-4.98$ & $0.50$\\
$\theta_2$ (gwesp(log(2)))           & $1.14$  & $0.17$\\
$\theta_3$ (practice - homophily)    & $0.63$ & $0.22$\\
$\theta_4$ (gender - homophily)      & $0.15$  & $0.22$\\
$\theta_5$ (practice - main effect)  & $0.20$  & $0.08$\\
\hline
\multicolumn{3}{c}{Between-model acc. rate: $0.04$}\\
\hline\hline
\end{tabular}
\caption{Summary of posterior parameter estimates and acceptance rates.}
\label{tab:lazega2_parpost}
\end{table}

The Bayes Factor for the comparison between model $m_2$ (best model) against model $m_1$ was around $2.32$ thus implying that there 
is not strong evidence to reject model $m_1$. From the results obtained above, we can state that the collaboration network is 
enhanced by the practice similarity effect. The first model highlights how the collaboration relations are strongly enhanced by 
having the same gender or practice. The positive value $\theta_2$ in both models indicates the presence of complex transitive 
effect captured by the edgewise shared partner statistics. 

\section{Discussion}
\label{sec:conclusions}

This paper has explored Bayesian model selection for posterior distributions with intractable likelihood functions. To our knowledge, this work represents a first step in the direction of conducting a Bayesian analysis of model uncertainty 
for this class of social network models. The methodological developments presented here have applicability beyond exponential random graph models, for example such methodology can be applied to Ising, potts or autologistic models. 

We introduced a novel method for Bayesian model selection for exponential random graph models which is based on a trans-dimensional extension of the exchange algorithm for exponential random graph models of \cite{cai:fri11}. 
This takes the form of an independence sampler making use of a parametric approximation of the posterior in order to overcome the issue of tuning the parameters of the jump proposal distributions and increase within-model acceptance rates.
We also note that the methodology may also find use in other recent papers which are also amenable to Bayesian analysis of networks such as \cite{kos:rob:pat10} for ERGMs in the presence of missing data and \cite{sch:han11} who implemented a version of the exchange algorithm adapted to hierarchical ERGMs with local dependence.

This methodology has been illustrated by four examples, and is reproducible using the {\tt Bergm} package for {\tt R} \citep{cai:fri12b}. Additionally we have presented a within-model approach for estimating the model evidence which relies on the path sampling approximation of the likelihood normalizing constant and nonparametric density estimation of the posterior distribution. 

The methods described in this paper have their limitations, however. The computational effort required by these algorithms render inference for large networks with hundreds of nodes or models with many parameters, out of range. Moreover, the need to take the final realisation from a finite run Markov chain as an approximate ``exact'' draw from the intractable likelihood is a practical and pragmatic approach. As yet a perfect sampling algorithm has not been developed for ERGMs, and this would have clear applicability for our algorithms.

% \paragraph{Acknowledgement} 
% Alberto Caimo was supported by an IRCSET Embark Initiative award and Nial Friel's research was supported by a Science Foundation Ireland Research Frontiers Program grant, 09/RFP/MTH2199.

\bibliographystyle{asa} %\bibliographystyle{macro/mybib} 
\bibliography{myref}

\end{document}